\documentclass[a4paper,11pt]{article}
\pdfoutput=1 

\usepackage{subcaption}
\usepackage{jcappub} 

\usepackage[T1]{fontenc} 
\usepackage{dsfont}
\def\be{\begin{equation}}
\def\ee{\end{equation}}
\def\bea{\begin{eqnarray}}
\def\eea{\end{eqnarray}}
\def\beq{\begin{eqnarray}}
\def\eeq{\end{eqnarray}}
\def\bas{\begin{subequations}\begin{eqnarray}}
\def\eas{\end{eqnarray}\end{subequations}}
\def\nn{\nonumber}

\def\lsim{\mathrel{\mathop  {\hbox{\lower0.5ex\hbox{$\sim$}
\kern-0.8em\lower-0.7ex\hbox{$<$}}}}}
\def\gsim{\mathrel{\mathop  {\hbox{\lower0.5ex\hbox{$\sim$}
\kern-0.8em\lower-0.7ex\hbox{$>$}}}}}

\def\be{\begin{equation}}
\def\ee{\end{equation}}
\def\bea{\begin{eqnarray}}
\def\eea{\end{eqnarray}}
\def\beq{\begin{eqnarray}}
\def\eeq{\end{eqnarray}}
\def\bas{\begin{subequations}\begin{eqnarray}}
\def\eas{\end{eqnarray}\end{subequations}}
\def\nn{\nonumber}

\newcommand{\cE}{{\mathcal E}}

\newcommand{\cH}{{\mathcal H}}
\newcommand{\cM}{{\mathcal M}}

\newcommand{\cR}{{\mathcal R}}

\newcommand{\cT}{{\mathcal T}}

\newcommand{\cS}{{\mathcal S}}

\def\rd{\textrm{d}}

\newcommand{\bes}{\begin{eqnarray}}
\newcommand{\ees}{\end{eqnarray}}

\def\nn{\nonumber}

\def\dd{\mathrm{d}}

\numberwithin{equation}{section}

\title{\boldmath Towards consistent black-to-white hole bounces from matter collapse}

\author[a,1]{J. Ben Achour,\note{Corresponding author.}}
\author[b]{S. Brahma,}
\author[a,c]{S. Mukohyama,}
\author[d,e]{J-P. Uzan,}

\affiliation[a]{Center for Gravitational Physics, Yukawa Institute for Theoretical Physics, Kyoto University, 606-85502, Kyoto, Japan}
\affiliation[b]{Department of Physics, McGill University, Montr\'eal, QC H3A 2T8, Canada}
\affiliation[c]{Kavli Institute for the Physics and Mathematics of the Universe (WPI), The University of Tokyo Institutes for Advanced Study, The University of Tokyo, Kashiwa, Chiba 277-8583, Japan}
\affiliation[d]{Institut d'Astrophysique de Paris, CNRS UMR 7095,
Universit\'e Pierre \& Marie Curie - Paris VI, 98 bis Bd Arago, 75014 Paris, France}
\affiliation[e]{Sorbonne Universit\'es, Institut Lagrange de Paris, 98 bis, Bd Arago, 75014 Paris, France}

\emailAdd{jibril.benachour@yukawa.kyoto-u-ac.jp, suddhasattwa.brahma@gmail.com, shinji.mukohyama@yukawa.kyoto-u-ac.jp, uzan@iap.fr}

\abstract{This article presents a new model-independent constraint for  bouncing black hole geometries. This constraint, applicable for a physical black hole formed by a collapsing compact object, sets a bound on the minimal allowed radius of the time-like surface of the collapsing star at the bounce. It follows from this bound that the shell always bounces in an untrapped region or precisely on a trapping horizon. This constraint is of purely kinematical origin as it descends from the continuity of the metric between the geometries describing the interior and exterior of the collapsing object so that it is completely model-independent. The second part of this work investigates the conditions  under which an effective extension of the Oppenheimer-Snyder collapse can describe consistent black-to-white hole bounces. Therefore, we present a general framework in which the collapse can lead to the formation of outer and inner horizons.  We show that on top of the previous kinematical constraint, an additional dynamical condition has to be satisfied in order for the model to admit well-defined black-to-white hole solutions. As expected, this second condition turns out to be model-dependent. The resulting class of models describing bouncing compact objects are characterized by three parameters for the star (mass, initial radius and density) and two quantum parameters descending from the UV-completion of the exterior and interior geometries. The solution space contains (1) bouncing stars and (2)  bouncing black holes and (3) a new class of astrophysical objects which alternate between a bouncing star and a bouncing black hole. Finally, we provide an explicit construction of a black-to-white hole bounce using the techniques of spatially closed LQC and discuss the novel properties induced by the presence of the inner horizon in this new framework. This generic framework lays down the foundation for interesting phenomenological investigations concerning the astrophysical properties of these objects, as well as a novel platform for further developments.}

\begin{document}

\begin{flushright} {\footnotesize YITP-20-46, IPMU20-0041}  \end{flushright}

\maketitle

\section{ Introduction} 
General Relativity (GR) predicts that for sufficiently dense matter, gravitational collapse can lead to the formation of a trapping horizon bounding a trapped region. Provided that the standard energy conditions are satisfied, the collapse inevitably generates pathological singularities which signal the breakdown of GR as a predictive theory at high energies. As such, the ultra compact objects detected so far through gravitational wave astronomy cannot be modeled by the standard black hole solutions of GR all the way down to the Planck scale. A complete description of such objects, therefore, requires an ultraviolet (UV) completion of GR.

While a full theory of quantum gravity is still out of reach, important lessons have however already been learnt from the quantization of symmetry reduced gravitational systems. The quantization of cosmological backgrounds result in a mixing of the classical contracting and expanding trajectories, providing bouncing cosmological dynamics. See Ref. \cite{Bojowald:2015iga} for a review. Similar results have also been obtained in the context of canonical quantum gravitational collapse. See Refs. \cite{Frolov:1979tu, Frolov:1981mz, Hajicek:1992mx, Hajicek:2000mi, Hajicek:2001yd, Liu:2014kra} for previous results in this direction and Refs. \cite{Kiefer:2019csi, Schmitz:2019jct, Piechocki:2020bfo} for more recent investigations. While the robustness of such bounces are still largely restricted to simple enough models and the covariance of the construction is yet to be demonstrated\footnote{In particular, in the context of LQC, the standard strategy of polymerization suffers from several drawbacks, among which a consistent implementation of covariance. See \cite{Bojowald:2020wuc, Bojowald:2019ujl, SpherCov, BlackCov} for more details.}, the idea that classically disconnected solutions of GR can be connected at the quantum level suggests an interesting possible outcome for gravitational collapse. As already argued in Ref.~\cite{Barcelo:2014npa}, the classical black and white hole solution sectors of gravitational collapse might also superpose at the quantum level to eventually leads to a black-to-white hole transition. This elegant idea raises many unanswered questions. What is the nature of the quantum transition when the geometry tunnels from a trapped to an anti-trapped region? Do the associated quantum gravity effects remain confined to the deep interior or can they extend up to the horizon scale? What is the time scale associated with this tunneling and how does it scale with the black hole mass for an asymptotic observer?  Is this mechanism a dominant or sub-dominant process w.r.t. Hawking evaporation? And, finally, can it lead to a resolution of the information loss paradox? 

These open questions can only be answered in the framework of a non-perturbative quantum theory of gravity where the transition amplitude associated to this phenomenon could be concretely computed\footnote{See Refs.~\cite{Christodoulou:2016vny, Christodoulou:2018ryl} for efforts in this direction.}. Such a theoretical framework being still out of reach despite numerous efforts, the strategy has been to investigate this plausible mechanism using effective methods; see Refs.~\cite{Carballo-Rubio:2018jzw, Malafarina:2017csn} for reviews. Several effective models of black-to-white hole transition have been constructed thus far, using different setups such as the black hole firework based on collapsing null shells \cite{Haggard:2014rza}, the shock wave approach~\cite{Barcelo:2014cla, Barcelo:2015uff, Barcelo:2016hgb}, and the different polymer black hole interior geometries inspired from LQG~\cite{Alesci:2019pbs, Bodendorfer:2019cyv, Ashtekar:2018cay}. See Refs.~\cite{DeLorenzo:2015gtx, Brahma:2018cgr, Rovelli:2018cbg, DAmbrosio:2018wgv, Bianchi:2018mml} for more detailed investigations of these different effective constructions and Refs.~\cite{Bodendorfer:2019xbp, Bojowald:2019dry, Bouhmadi-Lopez:2019hpp, Bodendorfer:2019nvy, Bodendorfer:2019jay, Assanioussi:2019twp, Cortez:2017alh, BenAchour:2018khr, Bojowald:2018xxu, Modesto:2009ve} regarding the LQG inspired models. Bouncing quantum black holes have also been discussed recently in \cite{Kiefer:2019csi, Schmitz:2019jct, Piechocki:2020bfo} using affine coherent state quantization. Since these models are based on very different assumptions, such as different choices of the scale at which quantum gravity effects become dominant, their conclusions remain very strongly model-dependent. In order to gain insight on some of the above questions, it is therefore crucial to develop as much as possible model-independent approaches which could serve to extract general properties of these possible bouncing compact objects. It has been shown that reversing this logic, and assuming that self consistent black holes are geodesically complete geometries, one can derive interesting generic constraints on the realization of such regular ultra compact objects even working within standard Riemannian geometry. See for example the recent classification of these geometries presented in Refs.~\cite{Carballo-Rubio:2019fnb,  Carballo-Rubio:2019nel}.  

More recently, the seminal Oppenheimer-Snyder-Datt (OSD) model of gravitational collapse introduced in Refs~\cite{Datt:1938, Oppenheimer:1939ue} was revisited along these same lines in Refs. \cite{BenAchour:2020bdt, BenAchour:2020mgu}. An interesting advantage of the OS model is to provide an simple platform to test the construction of bouncing compact object by importing techniques from quantum cosmology. Indeed, the classical OSD solution relies on the gluing of the Schwarzschild exterior geometry with the spatially closed Friedmann universe filled with dust, across a time-like thin shell which represents the surface of the star~\cite{Oppenheimer:1939ue}. Instead, by assuming the existence of a regular bouncing interior geometry, one can then investigate the consequences on the dynamics of the resulting bouncing compact object. Quite surprisingly, this exercise revealed the existence of model independent constraints on the allowed energy scale of the bounce. More precisely, it was shown in Ref. \cite{BenAchour:2020bdt} that if a bounce occurs, it can only occur above the threshold of horizon formation such that no trapped region can ever form. This conclusion holds independently of the mechanism responsible for the singularity resolution, and relies only on i) the continuity of the induced metric across the time-like surface of the star, and ii) the assumption of a single horizon exterior geometry (modelled by the classical Schwarzschild metric). This general construction was then used in Ref. \cite{BenAchour:2020mgu} to model a bouncing compact object using the effective bouncing dynamics of the spatially closed loop quantum cosmology (LQC) developed in Refs. \cite{Corichi:2011pg, Dupuy:2016upu}. 

The goal of the present work is to generalize the above results and investigate the consequences of this model-independent constraint found in Ref. \cite{BenAchour:2020bdt} when relaxing the assumption of the Schwarzschild exterior geometry, and allowing instead for a more general spherically symmetric exterior geometry with possibly multiple horizons. Indeed, the result obtained in Ref. \cite{BenAchour:2020bdt} suggested that in order to form a trapped region which eventually tunnels into an anti-trapped region, a missing key ingredient to include in the analyses was the formation of an inner horizon during the process. 

In order to test this idea and make the argument as general as possible, we will therefore relax in several ways the assumptions used in Ref. \cite{BenAchour:2020bdt} to derive our earlier result. To this end, we shall consider the gluing of two general spherically symmetric geometries across a time-like thin shell and only assume the existence of a bounce at some point in the dynamics. Using this rather general ansatz, we shall derive a generalization of our earlier result which explicitly relates the minimal allowed radius of the time-like surface of the collapsing object at the bounce with the Misner-Sharp mass of the exterior geometry. Since this quantity encodes the horizon structure, it allows one to discuss the consequences of having an inner horizon compared to the Schwarzschild case investigated in Ref. \cite{BenAchour:2020bdt}. This constraint at the bounce will now depend solely on the continuity of the induced metric across the time-like thin shell, and as such, will provide a kinematical constraint for such bouncing object, completely independent of the mechanism responsible for singularity-resolution. The major result descending from this constraint is that for such general gluing, the shell can bounce only in an untrapped region. Hence, while the formation of a trapped region was forbidden in Ref. \cite{BenAchour:2020bdt}, the more general setup discussed here allows one to describe a bouncing black hole. As anticipated in Ref. \cite{BenAchour:2020bdt}, the formation of an inner horizon turns out to be crucial in order to model such ultra compact objects.

The above kinematical constraint being only a necessary condition, additional assumptions are needed to build a well-defined model and a given dynamics has to be chosen through a choice of the second junction condition. We shall thus assume the standard Israel-Darmois matching conditions. Then, we will assume that the exterior geometry is static and possesses both an outer and an inner horizon, but the Misner-Sharp energy will be kept unspecified. The interior geometry will correspond to a bouncing spatially closed Friedmann-Lema\^{\i}tre (FL) universe filled with dust. Nevertheless, we shall also remain agnostic on the singularity resolution at play, and use the same parameterization of the quantum corrections as the one introduced in Ref.~\cite{BenAchour:2020bdt}. Using this generic setup, we present a general solution to the junction conditions and derive a generalized mass relation involving the Misner-Sharp mass function characterizing the exterior geometry, the energy content of the thin shell and the quantum corrections of the interior geometry. Starting from this new mass relation, we can extract a dynamical constraint which will play a crucial role in selecting the allowed dynamics. Interestingly, we will show that this constraint, although derived from a different perspective, encodes the same restriction on the energy scale of the bounce as the previous kinematical constraint derived in the first part. We will then deduce from it a classification of the allowed objects depending on the position of the minimal radius with respect to the outer and inner horizons.

To finish, we will use the lessons gained from deriving the constraints in the first two sections to construct an explicit example of a black-to-white hole bounce, for a collapsing star, within the thin shell formalism. In order to have a tractable exterior geometry describing a regular black hole with outer and inner horizons, we will use the recent proposal introduced in Ref. \cite{Simpson:2019mud}. The interior geometry of the collapsing matter will be modeled by the bouncing spatially closed LQC already used in Ref.  \cite{BenAchour:2020mgu}. Hence, both the exterior and interior geometries come with their own UV cut-off scales, denoted respectively $\beta$ and $\lambda$. Depending on the range of these two parameters, we will show that there exists well-defined solutions to the above constraints which allow one to describe an effective black-to-white hole bounce. The resulting bouncing compact object is therefore characterized by five parameters. The classical parameters correspond to its asymptotic ADM mass $M$ as well as its maximal radius $R_{\text{max}}$ and minimum energy density $\rho_{\text{min}}$. Contrary to the classical OS model, the density is now a free parameter and is not fixed by setting $(M, R_{\text{max}})$. Then, the quantum parameters are given by the UV cut-off $(\beta, \lambda)$ associated with the exterior and interior geometries. The consistency of the model imposes a precise hierarchy between these parameters. Interestingly, the existence of an inner horizon through the non-vanishing parameter $\beta$ allows one to apply the model to macroscopic objects, a possibility which was excluded in Ref. \cite{BenAchour:2020mgu} where only Planckian relics could be considered.  We need to emphasize that the thin shell turns out to be crucial in this construction as it encodes part of the quantum effects. As we shall see, the perfect fluid carried by the thin shell violates the weak energy condition only near the bounce but satisfies it otherwise.

To summarize, this work presents new model-independent constraints to construct effective black-to-white hole bounces in the context of the thin shell formalism. As opposed to eternal black holes, which are objects of purely mathematical interest, these constraints are applicable only for physical black holes being formed by matter collapse. In this  precise sense, our analyses goes much above and beyond previous works in which a black-to-white hole transition was considered only for vacuum spherically symmetric spacetimes, such as in LQG inspired models \cite{Alesci:2019pbs, Bodendorfer:2019cyv, Ashtekar:2018cay}. These new constraints underline the crucial role played by the formation of an inner horizon in the process, and suggest therefore new challenges for the model building of these bouncing black holes. While the explicit model we use as an example is effective and appears at best as illustrative, it nevertheless has the merit to provide a realization of a black-to-white hole bounce which encompasses several crucial ideas introduced in the initial Planck star proposal~\cite{Rovelli:2014cta}: it includes the role of collapsing matter, implements a scale of quantum gravity dictated by the density of the compact object, and regularizes the singularity using the techniques of LQC to finally describe a black-to-white hole transition where the quantum effects are confined in the deep interior region. The framework developed here should therefore serve as a starting point for additional investigations on not only the phenomenology but also the quantization of these bouncing black holes.

The article is organized as follows. Section~\ref{sec1} presents the general kinematical and model independent constraint for the construction of bouncing compact objects. Then, Section~\ref{sec2} presents the construction of the dynamics of such a compact object and introduces the dynamical constraint corresponding to the matching condition of the energy content in the exterior and interior regions. To finish, Section \ref{sec3} presents an explicit example of a black-to-white hole bounce.

\newpage
\section{Kinematical model-independent constraint for bouncing black holes}\label{sec1}

Let us begin by stating the explicit assumptions of our construction:
\begin{enumerate}
    \item We assume spherical symmetry in $(3+1)$-dimensions;
    \item We demand the continuity of the induced metric across a time-like thin shell representing the surface of a star joining two spherically symmetric (possibly dynamical) geometries;
    \item We assume the existence of a bounce in the interior geometry {as a resolution of the classical singularity}.
\end{enumerate}
In the following, we show that starting from this general set-up, one can extract a purely kinematical constraint on the energy scale of the assumed bounce and on the required horizon structure to form a consistent black-to-white hole bounce.

Let us consider two spherically symmetric spacetimes $\cM^{\pm}$ which describe, respectively, the exterior and the interior region of a compact object. Without any loss of generality, we can write these two geometries as
\begin{eqnarray}
\label{met}
\dd s^2_{\pm} = - e^{2\Phi_{\pm}(t_{\pm},r_{\pm})} f_{\pm}(t_{\pm},r_{\pm}) \dd t_{\pm}^2 + \frac{\dd r_{\pm}^2}{f_{\pm}(t_{\pm},r_{\pm})} + r_{\pm}^2 \dd\Omega^2\,,
\end{eqnarray}
where $(+)$ and $(-)$ correspond to the exterior and interior regions, respectively\footnote{Notice that we could have used the more general metric where $g^{\pm}_{\theta\theta}(t_{\pm},r_{\pm}) = R_{\pm}^2(t_{\pm},r_{\pm})$. However, except at the minimum of the physical radius $R(t,r)$, this metric can locally be written in the form (\ref{met}) using the Kodama foliation. Then, one can show that it does not modify the derivation of the kinematical constraint (\ref{condd}). }. In particular, notice that we allow here for fully dynamical geometries. It is a standard practice to parametrize the exterior and interior metric functions $f_{\pm}(t_{\pm},r_{\pm})$ as
\be\label{para}
f_{\pm}(t_{\pm},r_{\pm}) = 1 - \frac{2G m_{\pm}(t_{\pm},r_{\pm})}{r_{\pm}}\,,
\ee
where $m_{\pm}(t_{\pm},r_{\pm})$ are the Misner-Sharp masses of the exterior and interior regions, respectively, which are canonical scalar quantities associated to each regions. 

 
Consider then a time-like thin shell $\cT$ joining these two geometries. The unit spacelike vector, normal to this hypersurface, is denoted $n^{\mu} \partial_{\mu}$ and satisfies $n^{\mu}n_{\mu} = + 1$. We denote the intrinsic coordinates $\{ y^a\}$ on $\cT$ with $a \in \{ 0,1,2 \}$ and the four dimensional coordinates in $\cM^{\pm}$ as $\{ x^{\mu}_{\pm}\}$ with $\mu \in \{ 0,1,2,3 \}$.  The projection from the bulk regions $\cM^{\pm}$ to the time-like shell $\cT$ is performed via the projector $\; ^{\pm}e^{\mu}_a$ defined as
\be
\; ^{\pm}e^{\mu}_a = \frac{\partial x_{\pm}^{\mu}}{\partial y^a} \;, \qquad \; ^{\pm}e^{\mu}_a \; ^{\pm}e^{b}_{\mu} =  \delta^b_a\;, \qquad \; ^{\pm}e^{\mu}_a n_{\mu} =0
\ee 
The induced metric $\gamma^{\pm}_{ab}$ and the extrinsic curvature $K^{\pm}_{ab}$ of the hypersurface $\cT$ w.r.t. to each regions $\cM^{\pm}$ are then defined as
\be
\gamma^{\pm}_{ab}   = \; ^{\pm}e^a_{\mu} \; ^{\pm}e^{b}_{\nu} g^{\pm}_{\mu\nu} \;, \qquad    K^{\pm}_{ab} = \; ^{\pm}e^{\mu}_a \; ^{\pm} e^{\nu}_b \nabla_{\mu} n_{\nu}
\ee
where $\nabla$ is the covariant derivative compatible with the 4d metric $g^{\pm}_{\mu\nu}$ of the exterior/interior regions. In order to have a consistent gluing of the two geometries $\cM^{\pm}$ across $\cT$, one has to impose junction conditions. The set of conditions to be imposed descends from the theory of gravity under consideration, theory which admits $\cM^{\pm}$ as solutions of its field equations. In GR, these matching conditions are the well known Israel-Darmois junction conditions and can be split into a kinematical constraint imposing the continuity of the induced metric across $\cT$ and a second dynamical equation which equate the discontinuity of the extrinsic curvature across $\cT$ to the stress energy-momentum tensor of the shell. Conservation equations for this stress tensor follow from the Israel-Darmois junctions conditions and the ADM and scalar constraint evaluated on the shell. When considering higher order gravity theories, these junctions conditions are supplemented with additional dynamical conditions involving the additional degrees of freedom\footnote{See \cite{Senovilla:2013vra} for the junctions of $f(\cR)$ theory for example.}. Nevertheless, the continuity of the induced metric across $\cT$ is shared by any covariant theories of gravity. As our goal is to derive a model-independent constraint for bouncing compact objects, a natural strategy is to focus on this kinematical condition and check what are the consequences when assuming a bouncing interior dynamics. Therefore, in the following, we shall focus on the consequences of only imposing 
\be
[\gamma_{ab}] = 0
\ee
where the bracket of a quantity $X$ corresponds to $[X] = \left( X_{+} - X_{-} \right)|_{\cT}$. 

Let us denotes the coordinates $\{ y^a\}$ on $\cT$ as $\{ y^a\} = \{\tau, \theta, \phi\}$ such that $\tau$ is the proper time of the shell and the angular coordinates on $\cT$ and on $\cM^{\pm}$ are identified.  Hence, the time and radial coordinates in the exterior and interior regions can be parametrized as $t_{\pm} = T_{\pm}(\tau)$ and $r_{\pm} = R_{\pm}(\tau)$ on $\cT$ such that the induced metric on $\cT$ reads
\begin{eqnarray}
\dd s^2_{\pm} \big{|}_{\cT} & = &-  \left[ e^{2\Phi(T_{\pm},R_{\pm})} f(T_{\pm},R_{\pm}) \dot{T}_{\pm}^2 - \frac{\dot{R}_{\pm}^2}{f(T_{\pm},R_{\pm})} \right] \dd\tau^2 + R_{\pm}^2(\tau) \dd\Omega^2\,,
\end{eqnarray}
where a dot corresponds to a derivative w.r.t. the proper time  $\tau$. Therefore, the induced metric on $\cT$ can be rewritten as
\begin{eqnarray}
\dd s^2_{\pm} \big{|}_{\cT}= - \dd \tau^2+ R^2(\tau) \dd\Omega^2\,.
\end{eqnarray}
The continuity of the induced metric across $\cT$ imposes the following key relations
\begin{eqnarray}
\label{keycond}
e^{2\Phi(T_{\pm},R_{\pm})} f(T_{\pm},R_{\pm}) \dot{T}_{\pm}^2 - \frac{\dot{R}_{\pm}^2}{f(T_{\pm},R_{\pm})} = 1\,, \qquad R_{+}(\tau) = R_{-}(\tau)\,,
\end{eqnarray}
which are evaluated on the shell $\cT$. The second condition allows us to drop the subscript for the radius of the shell and denote it simply as $R(\tau)$. Let us again stress that these conditions are purely kinemtical and valid for any metric theories of gravity. 

Let us now focus on the function $A_{+}(\tau) = \dot{T}^{-1}_{+}$ associated to an observer on $\cT$ with the exterior time coordinate $T_{+}$. From Eq.~(\ref{keycond}), one obtains
\begin{eqnarray}
\label{A}
A^2_{+} (\tau)= e^{2\Phi_{+}(\tau)} f_{+}^2(\tau) \left[ f_{+}(\tau) + \left( \frac{\dd R(\tau)}{\dd\tau} \right)^2\right]^{-1}.
\end{eqnarray}
If there is a bounce, as we assumed, then $R(\tau)$ reaches a minimum $R_{b}$ at the bounce time $\tau_{b}$, i.e.,
\begin{eqnarray}
\dot{R}(\tau) \big{|}_{\tau_{b}}=0\;, \qquad \ddot{R}(\tau) \big{|}_{\tau_{b}} >0\,.
\end{eqnarray}
Then, at the bounce, the lapse function (\ref{A}) satisfies
\begin{eqnarray}
\label{atb}
A_{+}^2(\tau) \big{|}_{\tau_{b}}= e^{2\Phi_{+}(\tau)} f_{+}(\tau) > 0\,.
\end{eqnarray}
Using the parameterization (\ref{para}), the Lorentzian signature of the metric implies
that $e^{2\Phi_{+} (\tau)}>0$
 everywhere and that $f_{+}(\tau) > 0$ at $\tau_{b}$, which means that
\begin{eqnarray}\label{condd}
R\left(\tau_{b}\right) \geqslant 2 m_{+} \left[T_{+}\left(\tau_{b}\right), R\left(\tau_{b}\right)\right]\,.
\end{eqnarray}
This inequality constitutes our main result for this section. It sets a lower bound on the minimal radius that can be reached by the time-like surface of the collapsing star. Note that all our conclusions so far only  rely on the continuity of the induced metric across the thin-shell and is valid for any spherically symmetric and possibly dynamical geometries. This bound on the radius can equivalently be translated to an upper bound on the maximum energy scale at which the bounce can take place. To illustrate this further, it is useful to compute the expansion of out-going and in-going null rays evaluated on $\cT$. While their individual expressions depend on the choice of null vectors, a canonical quantity is given by their product. Provided the foliating $2$-surface is spherically symmetric, the product of the expansions for any pairs of in-going and out-going nulls vectors is given by
\begin{eqnarray}
\theta_{>} \theta_{<} (\tau)= \frac{4 }{R^2(\tau)} \left\lbrace\frac{2M_{+} \left[T_{+}\left(\tau\right),R\left(\tau\right)\right]}{R\left(\tau\right)} - 1\right\rbrace\,.
\end{eqnarray}
where the subscript $>$ (resp. $<$) corresponds to out-going (resp. in-going) null rays. 
The constraint (\ref{condd}) implies that 
\begin{eqnarray}\label{e.14}
\theta_{>} \theta_{<} (\tau_b) \leqslant 0\,,
\end{eqnarray}
This means that the hyper-surface $\cT$ corresponding to the surface of the compact object can bounce only in an untrapped region in which $\theta_{>} \theta_{<} (\tau_b)  <  0$ or precisely on a trapping horizon for which $\theta_{>} \theta_{<} (\tau_b) = 0$. The outcome of the collapse, therefore, depends on the horizon structure of the exterior geometry.

To illustrate the kinematical condition \eqref{condd}, consider the following two choices of the Misner-Sharp mass function.
\begin{enumerate}
\item The Schwarzschild spacetime  has $m_{+} [T_{+}(\tau), R(\tau)] : = M$ for all $\tau$. The condition~\eqref{condd} simply  implies that the compact object can never cross its Schwarzschild radius so that the bounce must occur outside this region, i.e.  $R(\tau_b) \geqslant 2GM$. This obviously prevents the formation of any trapped region. This confirms the result we obtained in Ref.~\cite{BenAchour:2020bdt}. 
\item If the mass function is such that $f_{+}$ has two zeros, which encode the locations of an outer and an inner horizon, the dynamical radius of the shell can in principle cross the outer horizon and the collapse can lead to the formation of a trapped region followed by the formation of an inner horizon. However, the condition~(\ref{e.14}) implies that the shell has to bounce in an untrapped region where $\theta_{>} \theta_{<} (\tau_b) < 0$ or on a trapping horizon where $\theta_{>} \theta_{<} (\tau_b) = 0$. In the former case, a bouncing black hole can exist only if an inner horizon is formed in the process\footnote{If the bounce happens when $\theta_{>} \theta_{<} (\tau_b) = 0$, either the bounce coincides with the formation of the inner horizon, or it corresponds to the formation of more exotic transition hyper-surface where $\theta_{>} (\tau_b) =0$ and $\theta_{<} (\tau_b) = 0$ simultaneously. This second case is realized in LQG inspired models of vacuum spherical symmetry such as \cite{Bodendorfer:2019cyv, Ashtekar:2018cay}, but is inconsistent with the dynamical constraint considered later on when the interior is modeled with a spatially closed FL universe filled with dust. See the discussion below Eq~(\ref{e.kodama}).}. It is obvious that if one wants to keep the quantum corrections confined to a small region in the deep interior, we would have to consider this case with the mass function having multiple zeros, underlining the crucial role palyed by the inner horizon in this process.
\end{enumerate}
Before going any further, let us emphasize that the constraints (\ref{condd}) and (\ref{e.14}) are purely kinematical and descend solely from {\em i)} imposing the continuity of the induced metric on the time-like thin shell $\Sigma$ and {\em ii)} assuming the UV-completion of the matter collapse model via the existence of a bounce in the dynamics of the interior geometry. In order to show the robustness of our result, it is useful to reproduce the derivation of this kinematical constraint in a gauge where the metric is not singular at the horizon. This has been done in the Appendix using instead the Gullstrand-Painlev\'e coordinates for a generally spherically symmetric spacetime.


\section{From the kinematical to the dynamical constraint}\label{sec2}

Up to now, we have obtained a general kinematical bound on the energy scale of the bounce, but this condition is not sufficient and one has to show that the shell can indeed cross the outer horizon and form a trapped region (which can eventually turn into an anti-trapped region) in order to form a bouncing black hole. While the previous result is completely model-independent, this second condition will largely depend on the model under consideration. Three elements have to be specified: the geometry of the outer spacetime, i.e. the function $m(t,r)$ in Eq.~(\ref{para}), the form of the quantum corrections and the set of junction conditions to be imposed.

\subsection{Parametrization of the exterior and interior geometries}

We start by specifying the exterior and interior geometries.

\subsubsection{Interior spacetime} Following our previous works~\cite{BenAchour:2020bdt, BenAchour:2020mgu}, we describe the inside of the collapsing star by a homogeneous Friedmann-Lema\^{\i}tre spacetime with spherical spatial sections,
\be
\dd s^2 = - \dd\tau^2 + R^2_c a^2(\tau) \left( \dd\chi^2 + \sin^2{\chi} \dd\Omega^2 \right)\,.
\ee
where $a(\tau)$ is the scale factor, $\tau$ the proper time of the interior region and $1/R^2_c$ corresponds to the constant spatial curvature of the geometry.
Its dynamics is dictated by the quantum cosmology equations for a pressureless dust, through the function $\Psi_1$ entering the modified Friedmann equation,
\be
\label{para1}
H^2 = \left(\frac{8\pi G}{3}\rho -\frac{1}{R_c^2a^2}\right)\left[1-\Psi_1(a)\right]\,.
\ee
The second Friedman equation can be parametrized similarly and the quantum corrections have to be related by a consistency condition. See Ref. \cite{BenAchour:2020mgu} for more details.  We assume that $\Psi_1$ is such that there exists a minimum value of the radius of the star, i.e. a value $a_{\rm min}$ such that
\be
\Psi_1(a_{\rm min})=1\,, \qquad 0<a_{\rm min}<1\,,
\ee
and that
\be
 0\leq 1-\Psi(a)\,, \leq1\quad \forall a\in [a_{\rm min},1]\,.
\ee
From the conservation of the stress-energy tensor, the energy density of the dust scales as 
\be
\label{ene}
\rho (\tau) = \frac{\cE}{a^3(\tau)} \,, \qquad \text{with} \qquad \cE = \frac{3}{8\pi G R^2_c}\,
\ee 
such that $\cE$ corresponds to the minimal energy density of the star prior to collapse. Hence, $R_c$  actually encodes the minimal energy density of the compact object. Notice that in general, we do not expect the continuity equation to keep its classical form in such effective quantum corrected cosmology. However, we shall use this simplification for two reasons. First, this scenario is realized in the case of spatially closed loop quantum cosmology, which we shall use in the third section in order to model our interior bouncing geometry. Moreover, it can be shown that several results obtained in the next sections are either independent of this modification, or only affected quantitatively. Therefore, we shall work with this simplification for now on.

The induced metric of the shell $\cT$ w.r.t the interior region is given by
\be
\label{intmet}
{\rm d}s^2_{-} = \gamma^{-}_{ab} {\rm d}y^a {\rm d}y^b = - {\rm d}\tau^2 + R^2_c a^2(\tau) \sin^2{\chi_0} {\rm d}\Omega^2
\ee
This amounts to choosing that the proper time of the shell and the proper time of the homogeneous interior region coincide. Then the shell is located at $\chi := \chi_0$ and the unit normal co-vector to the shell is given by $n_{\mu}dx^{\mu} = R_c a(\tau) d\chi$. This choice is not unique and one could work with a more general set up where the normal vector is not aligned with the direction $d\chi$. However, this setting simplifies several computations and allows one to avoid non-vanishing gradient of the stress energy tensor on $\cT$ when performing the gluing. The extrinsic curvature is easily computed and is given by
\be
\label{ExCurv1}
K^{-}_{ab} {\rm d} y^a {\rm d} y^b = R_c a(\tau) \sin(\chi_0) \cos(\chi_0) {\rm d}\Omega^2
\ee
The above embedding of the thin shell w.r.t the interior geometry implies that the outer radius of the star is
\be
R(\tau) = R_{\max} a(\tau)\,, \qquad R_{\text{max}} = R_c \sin{\chi_0}\,,
\ee
with the initial conditions at $\tau=0$ given by $a(0) =1$ and $\dot{a} =0$, so that $R_{\text{max}}$ is the maximal radius of the star.
This concludes the presentation of the interior geometry $\cM^{-}$ and its relation to the thin shell $\cT$.

\subsubsection{Exterior spacetime}  

In order to simplify the problem, we shall restrict the exterior geometry to be static and spherically symmetric 
\begin{eqnarray}
\label{eemet}
\dd s^2_{+} = -  f(r) \dd t^2 + \frac{\dd r^2}{f(r)} + r^2 \dd\Omega^2\,, 
\end{eqnarray}
with
\be\label{e.15}
f(r) = 1 - \frac{2Gm(r)}{r} = 1- \frac{2G M}{r} \; k\left(\frac{2GM}{r}\right)
\ee
such that $M$ is a constant and the function  $k$ behaves as\be
k\left(\frac{2GM}{r}\right) \rightarrow1 \qquad \text{for} \qquad r\rightarrow+\infty
\ee
so as to recover classical Schwarzschild far away from the star (for a star of mass $M$). We require that the spacetime has two horizons, so that $k$ is such that $f(r)$ has two zeros $R_{\text{out}}$ and $R_{\text{in}}$, i.e.
\be
 \frac{2GM}{R_{\text{out/in}}} k\left(\frac{2GM}{R_{\text{out/in}}} \right) = 1 \,, \qquad 0<R_{\text{in}}<R_{\text{out}}\,.
\ee
Let us point that assuming the exterior geometry is singularity free and possesses an inner horizon at $R_{\text{in}}$ implies that it is characterized by an additional parameter which encodes the deviation from the vacuum Schwarzschild geometry. As such, it plays the role of a UV cut-off in the exterior geometry. We shall denote this parameter $\beta$. It will be used explicitly when working with the specific choice (\ref{e.15b}) for the function $k$ in Section~\ref{sec4.2}.

The proper time of the shell being denoted $\tau$, the exterior and interior time coordinates can be parametrized as $t:= T(\tau)$ and $r:= R(\tau)$ on the shell $\cT$. Then, the four dimensional extension of the three dimensional tangent time-like unit vector $u^a \partial_a = \partial_{\tau}$ associated to an observer at rest on $\cT$ is given by $u^{\mu}\partial_{\mu} = e^{\mu}_a u^a \partial_{\mu} $ and reads explicitly
\be
u^{\mu}\partial_{\mu} = \dot{T} \partial_T + \dot{R} \partial_R  \;, \qquad u_{\mu} {\rm d}x^{\mu} = - f \dot{T} {\rm d} T + \frac{\dot{R}}{f} {\rm d} R
\ee
such that $u^{\mu} u_{\mu} = -1$ and where a dot corresponds to a derivative w.r.t $\tau$. The normal spacelike vector $n^{\mu} \partial_{\mu}$ to $\cT$ is given by
\be
n^{\mu}\partial_{\mu} = \frac{\dot{R}}{f} \partial_T + f \dot{T} \partial_R \;, \qquad n_{\mu} {\rm d} x^{\mu} = - \dot{R} {\rm d} T + \dot{T} {\rm d} R
\ee
such that $n^{\mu} u_{\mu} =0$. 
The induced metric on $\cT$ is given by
\be
\label{extmet}
{\rm d}s^2 = \gamma^{+}_{ab} {\rm d} y^a {\rm d}y^b = -  \left[ f \dot{T}_{\pm}^2 - \frac{\dot{R}^2}{f} \right] \dd\tau^2 + R^2(\tau) \dd\Omega^2
\ee
where the angular coordinates on $\cT$ and $\cM^{+}$ have been identified. Using the normal vector to $\cT$, the extrinsic curvature of $\cT$ w.r.t. $\cM^{+}$ is easily computed and reads
\be
\label{ExCurv2}
K^{+}_{ab} {\rm d} y^a {\rm d} y^b = - \frac{1}{ f \dot{T}} \left[ \ddot{R}   + \frac{f_{,R}}{2} \right] {\rm d} \tau^2 +  \dot{T} f R {\rm d} \Omega^2
\ee 
Defining the associated scale factors by $R_{\text{out/in}}=R_c\,\sin\chi_0\, a_{\text{out/in}}$ and evaluating the function $k$ on the shell $\cT$, we can write
\be
\frac{a_*}{a_{\text{out/in}}} k\left(\frac{a_*}{a_{\text{out/in}}} \right)=1\,, \qquad 0<a_{\text{in}}<a_{\text{out}} <1\,,
\ee
where we have introduced the two parameters
\be
\label{papa}
\sin\chi_s \equiv\frac{2GM}{R_c}\,, \qquad
a_*\equiv\frac{\sin\chi_s}{\sin\chi_0}\,.
\ee
Since the star is initially not a black hole, we have $0<a_*<1$.  The exterior geometry depends thus on the three parameters $(\chi_0,\chi_s,\cE)$ describing the star and also on the additional parameter encoding the existence of the inner horizon (and thus the deviation from Schwarzschild) which can be understood as a UV cut-off.
This concludes the presentation of the exterior geometry $\cM^{+}$ and its relation to the thin shell $\cT$.

\subsection{Junctions conditions}

We now need to specify the junctions conditions to be imposed on the shell. In principle, this junction conditions shall be derived from the UV complete theory of gravity under consideration, which admits both the exterior and interior geometries introduced above as solutions of its effective field equations. However, in practice, the derivation of these junctions conditions can be highly complicated if not impossible. For example, if one is interested in performing the gluing of LQG inspired geometries, such derivation is out of reach. In order to make progress, it is therefore tempting to use the standard Israel-Darmois (ID) junctions conditions as a first step and adopt the following point of view. Both exterior and interior solutions can be regarded as being GR solutions associated to an exotic energy-momentum tensor in the exterior and interior regions. In particular, the energy-momentum tensor violate the standard energy conditions such that it provides regular geometries both for the exterior and interior of the compact object. From this point of view, one can then introduced the ID junction conditions for a time-like shell which read
\begin{align}
\label{Kin}
[\gamma_{ab}] & = 0 \\
\label{Dyn}
[K_{ab} - \gamma_{ab} K] &= - 8\pi G \cS_{ab}
\end{align}
where $\cS_{ab}$ is the stress energy-momentum tensor on the time-like shell $\cT$ and where the bracket of a quantity $X$ corresponds to the discontinuity of this quantity across $\cT$. In the first section, we have discussed in details the consequences of the kinematical condition (\ref{Kin}). In the following, we shall investigate the consequences of the additional dynamical constraint (\ref{Dyn}). We recall that the above junctions conditions together with the vectorial and scalar constraints of General Relativity imposed on $\cT$ allow one to derive two additional tensorial conservation equations for the shell. However, these equations being a consequence of the junction conditions and the Einstein equations in the bulk, they do not add any new information. See \cite{D,WI} for more details.

As first step, let us write down the constraints imposed by the kinematical condition (\ref{Kin}). Using Eq.~(\ref{extmet}) and Eq.~(\ref{intmet}), we obtain
\be
\label{kinmacthing}
f \dot{T}_{\pm}^2 - \frac{\dot{R}^2}{f} = 1 \;, \qquad R(\tau) = R_c a(\tau) \sin{\chi_0}
\ee
The consequence of the first condition has been discussed in detail in the first section. In particular, one obtains that the differential of the exterior and interior time coordinate are related through
\be
\label{lapse}
\left(\frac{\rd T}{\rd\tau} \right)^{-2}= f^2 \left[ f+ \left( \frac{\dd R}{\dd\tau} \right)^2\right]^{-1}
\ee
As a second step, we write down the dynamical junction condition (\ref{Dyn}). The stress energy-momentum tensor $\cS_{ab}$ on $\cT$ is assumed to be a perfect fluid such that
\be
\cS_{ab} = \sigma u_a u_b + p \left( \gamma_{ab} + u_a u_b \right)
\ee
where $u_a = \left( 1,0,0\right)$ is the $3$-velocity of the observer at rest with the fluid. Then, one has that $\cS^a{}_b = \text{diag}\left( -\sigma, p, p\right)$.
This surface stress tensor is allowed to violate the standard energy conditions. In order to write the different components of the tensorial equation (\ref{Dyn}) in a compact form, we introduce the notation $\kappa^a{}_b = [K^a{}_b - \delta^a_b K]$. Then, one has
\begin{align}
\sigma = - \frac{\kappa^{\theta}{}_{\theta}}{4\pi G} \;, \qquad p = \frac{1}{8\pi G} \left( \kappa^{\tau}{}_{\tau} + \kappa^{\theta}{}_{\theta}\right)
\end{align}
Using the components of the extrinsic curvature given by Eq.~(\ref{ExCurv1}) and Eq.~(\ref{ExCurv2}), one obtains that
\begin{align}
\label{en}
\sigma & = \frac{1}{4\pi G R} \left( \cos{\chi_0} - f \dot{T} \right) \\
\label{press}
\sigma + 2p & =  \frac{1}{4\pi G  f \dot{T}} \left[ \ddot{R}   + \frac{f_{,R}}{2} \right] 
\end{align}
This fixes the time-evolution of the surface energy and surface pressure during the whole bouncing dynamics. We shall now analyze further the consequences of the equation (\ref{en}) on the dynamics of the bouncing compact object.

\subsection{Mass relation and the dynamical matching constraint}

The relation (\ref{en}) fixing the profile of the stress energy on $\cT$ is the key relation to derive the standard mass relation in the Oppenheimer-Snyder model where $\sigma=0$. Below, we investigate the modification to the classical mass relation induced by our choice of exotic regular geometries. 

Using the formula (\ref{lapse}) together with the modified Friedman equation (\ref{para1}) and the parametrization of the metric function $f$ in term of the Misner-Sharp mass $m$, Eq~(\ref{en}) can be recast into 
\be
\label{Master-Equation}
m = \frac{4\pi}{3} \rho R^3 - \frac{R^3}{2 G} \left[ \Sigma^2 - \frac{2 \cos{\chi_0}}{R} \Sigma+ \left( \frac{8\pi G}{3} \rho - \frac{1}{R^2_c a^2}\right) \Psi_1 \right]
\ee
where we have introduced the notation $\Sigma = 4\pi G \sigma$.  It provides the generalized mass relation for our bouncing compact object. As expected, when the energy of the shell vanishes, i.e $\Sigma=0$ and the quantum correction are ignored $\Psi_1 =0$, one recovers the Oppenheimer-Snyder mass relation. Finally, using the formulas (\ref{e.15}) and (\ref{ene}), one can rewrite this key relation as 
\begin{align}
\label{seconorder}
\frac{a_*}{\sin^2\chi_0} k\left(\frac{a_*}{a} \right) & =  1-  a^3\left[ \tilde{\Sigma}^2 - \frac{2\cos{\chi_0 }}{a\sin\chi_0} \tilde{\Sigma} +  \frac{1 -a}{a^3} \Psi_1 \right],
\end{align}
where we have introduced $\tilde{\Sigma} = R_c \Sigma$ such that $[\tilde{\Sigma}] = 1$.
Interestingly, this mass relation takes the form of a second order polynomial equation for  $\Sigma$. It will  be well-defined only if the determinant of Eq.~(\ref{Master-Equation}) remains positive on $[a_{\rm min},1]$, i.e. if
\begin{align}\label{conddd}
\Delta = \frac{4}{a^2 \sin^2{\chi_0}} \left[  1-\frac{a_*}{a}k\left(\frac{a_*}{a}\right) +\sin^2\chi_0\left(\frac{1}{a}-1\right)(1-\Psi_1(a)) \right]  \geqslant 0\,.
\end{align}
This relation relates only the Misner-Sharp energy, i.e. $k(a)$, and the quantum correction $\Psi_1(a)$. Remarkably, one can show that this dynamical constraint  actually contains the kinematical condition obtained in the first section. Indeed, using that at $a=a_{\text{min}}$, one has $\Psi_1(a_{\text{min}})=1$, and one obtains
\be
1-\frac{a_*}{a_{\text{min}}}k\left(\frac{a_*}{a_{\text{min}}}\right) \geqslant 0\,.
\ee
Then, upon using Eqs.~(\ref{e.15}) and (\ref{papa}), this inequality at the bounce coincides exactly with the kinematical condition~(\ref{condd}) derived in the first section\footnote{Interestingly, the assumption that the continuity equation is classical is not necessary to re-derive the kinematical constraint from the dynamical one, showing further evidence for the generality of this result.}. Finally, notice that if $k(a)=1$ (i.e. the Schwarzschild geometry) and assuming no quantum corrections i.e. $\Psi_1 =0$, then $\Sigma=0$ leads to the standard OS mass relation. 

\begin{figure}[h]
 \centering
   \includegraphics[scale= .4]{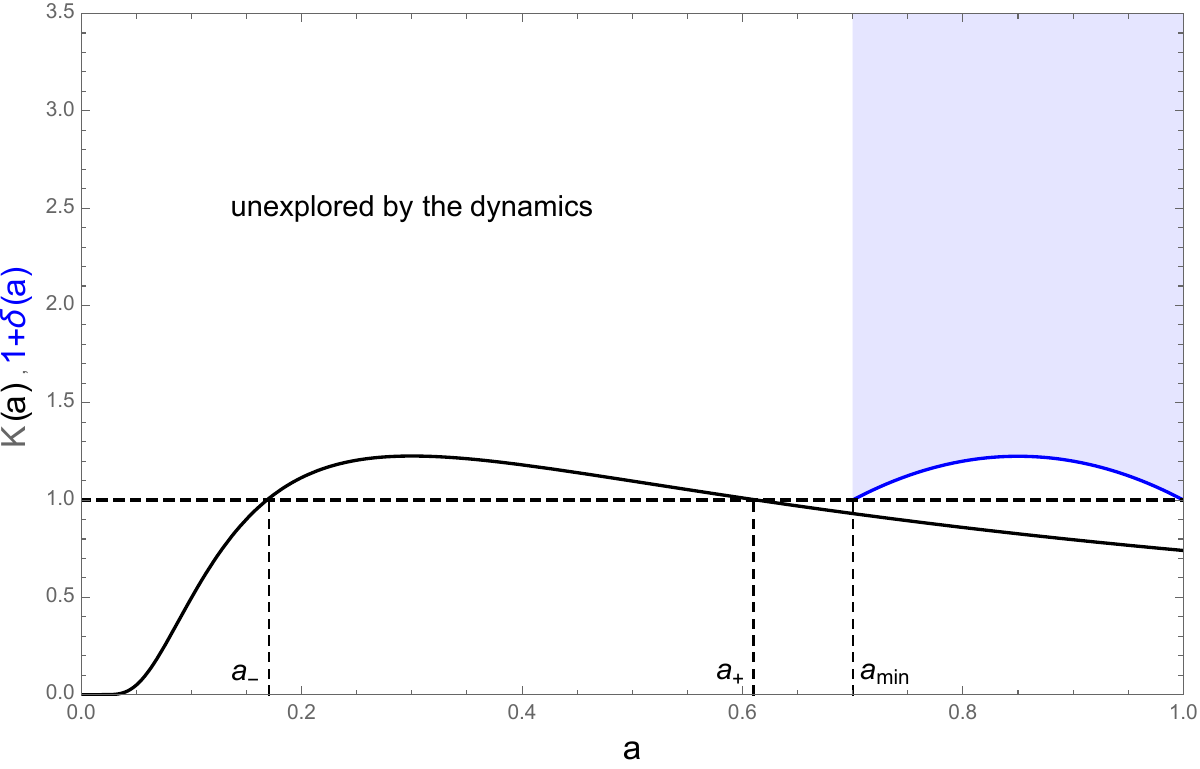} \includegraphics[scale= .4]{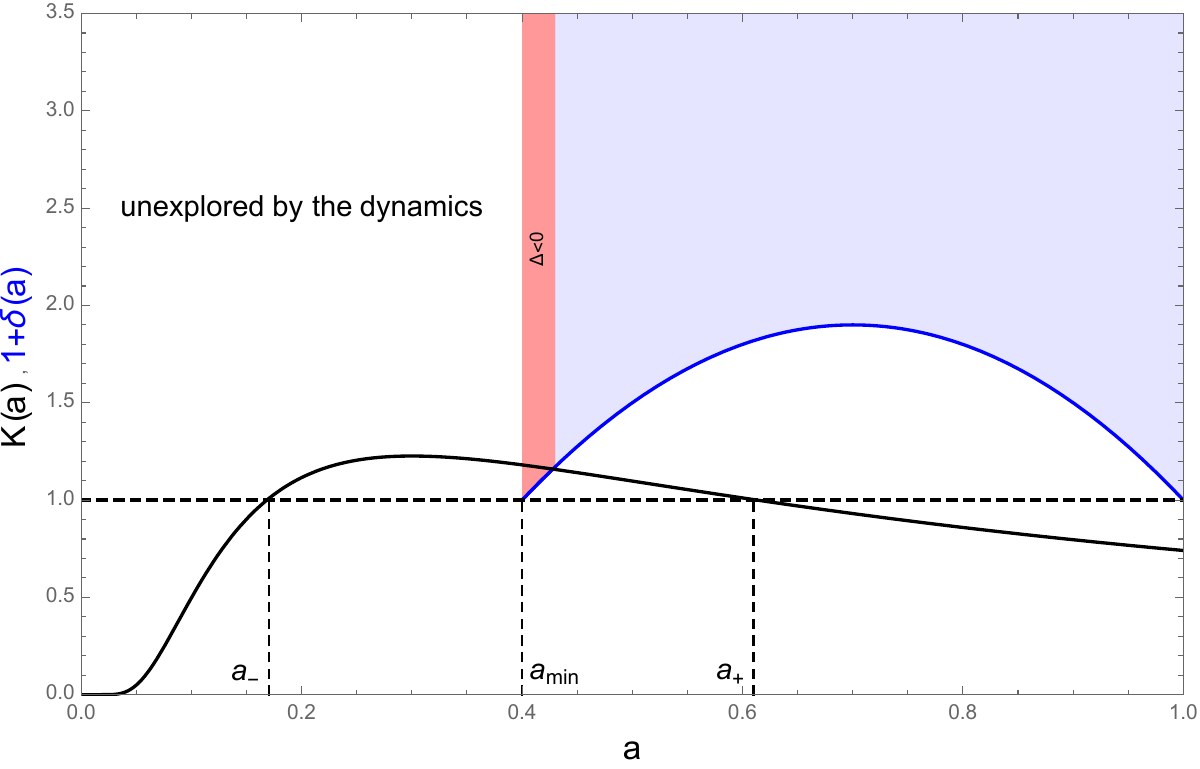} \includegraphics[scale= .4]{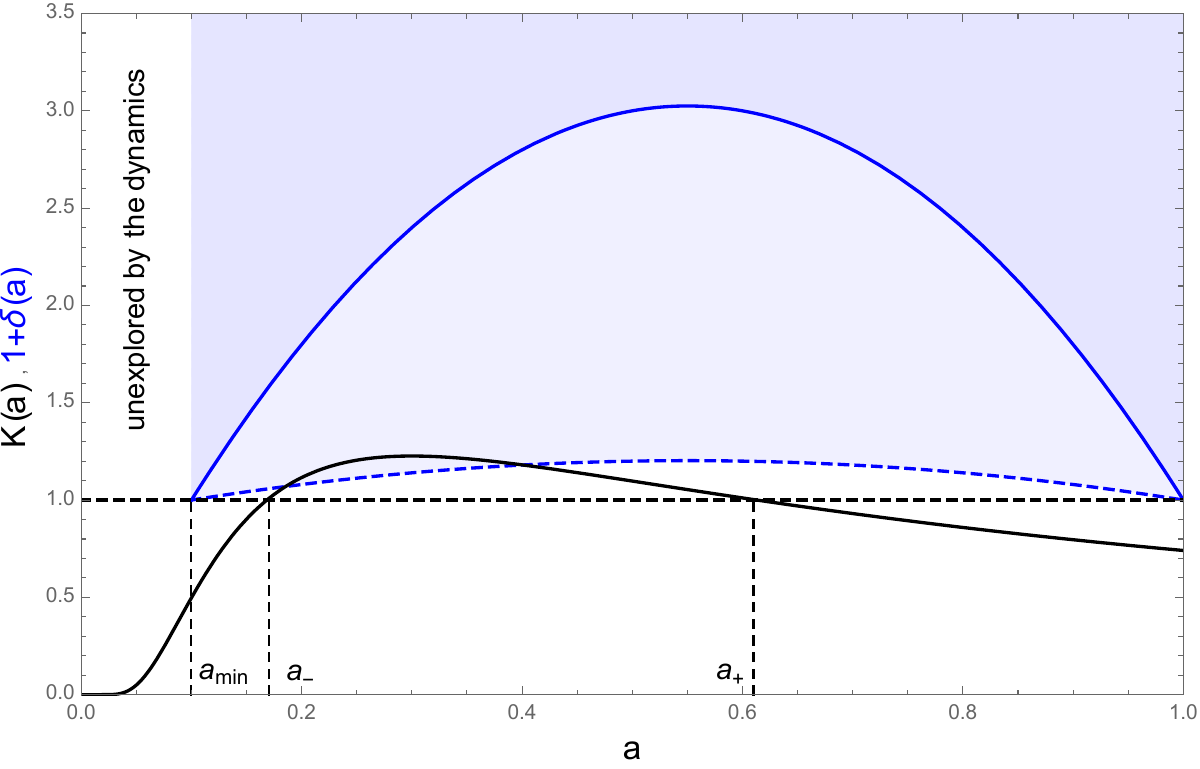}
   \caption{The three possible dynamics of a bouncing star. On each plot the black curve depicts $K(a)$ and the blue curve $1+\delta(a)$. The allowed range of scale factor for the dynamics in $[a_{\rm min},1]$ and $K(a)$ shall not cross the shaded region for the model to be well-defined. {\em Up}- the star bounces before crossing the outer horizon. Such a model provides a model for a bouncing star. {\em Middle}- when $a_-<a_{\rm min}a_+$ the two curves necessary cross on $[a_{\rm min},a_+[$ and the surface energy density is ill-defined on the red zone where the condition~(\ref{conddd}) is violated. Note that a similar situation arises if $a_+>1$. {\em Bottom}- when $a_{\rm min}<a_-$ the conclusion will depend on the specific form of $k(a)$ and $\Psi_1(a)$. If  the curves cross (dashed case for $1+\delta(a)$) the model is ill-defined but there exist configurations (solid line) in which the model remain well-defined while the star crosses both the outer and inner horizons. This would provide a dynamical model of a bouncing black hole.}
   \label{fig0} 
\end{figure}

To go further, let us introduce the notations
\begin{eqnarray}
K(a)\equiv\frac{a_*}{a}k\left(\frac{a_*}{a}\right), \qquad
\delta(a)\equiv \sin^2\chi_0\left(\frac{1}{a}-1\right)[1-\Psi_1(a)].
\end{eqnarray}
It follows from our hypothesis that $\delta(1)=\delta(a_{\rm min}) =0$ and that $1+\delta(a)\geq1$ on $[a_{\rm min},1]$. Then the function $K$ is such that $K(a\rightarrow+\infty)=0$ and $K(a_+)=K(a_{-})=1$. We also assume that $K(a) <1$ for $a < a_{-}$, corresponding to the untrapped region inside the inner horizon. Then, the condition~(\ref{conddd}) is rewritten as 
\be\label{e.6b}
K(a)\leq  1+\delta(a)\,.
\ee
This inequality leads to only three possible configurations, that are summarized in Fig.~\ref{fig0}:
\begin{enumerate}
\item\underline{If $a_{\rm min}\in ]a_+,1[$}, the star bounces before reaching the outer horizon. Since $a_{\rm min}>a_-$, we have 
\be
 a_{\rm min} < K(a) < 1\,, \qquad  a_{\rm min} < 1 < 1 + \delta(a)\,,
\ee
so that the condition~(\ref{e.6b}) is fulfilled. This is depicted in Fig.~\ref{fig0}-{\em left}. Examples of such configurations were given in Ref.~\cite{BenAchour:2020mgu}.
\item\underline{If $a_{\rm min}\in ]a_-,a_+[$}, then 
\be
K(a_{\rm min})> 1+\delta(a_{\rm min}) = 1\,,
\ee 
and the condition~(\ref{e.6b}) will be violated before reaching the bounce. This is depicted by the red zone on Fig.~\ref{fig0}-{\em middle} which arises from the fact that the two curves $ 1+\delta(a)$ and $K(a)$ need to cross. Hence, the bounce cannot occur between the two horizons. This is compatible with our general constraint~(\ref{e.14}). Note however that this same conclusion is derived here from the dynamical properties, i.e. from the second junction condition on the extrinsic curvature, and not from the first (non dynamical) junction condition on the induced metric which was not needed in this case, and was {\em sufficient} to prove Eq.~(\ref{e.14}).
\item\underline{If $a_{\rm min}\in ]0,a_-[$} the conclusion on the viability of the model will depend on the explicit form of $k(a)$ and $\Psi_1(a)$ but note  that the sufficient requirement will be that
\be\label{e.ccc}
K(a)<1+\delta(a)\,,\quad \hbox{on}\quad ]a_-,a_+[\,,
\ee 
for the model to be compatible with the condition~(\ref{e.6b}) since $K(a) < 1$ for $a < a_{-}$ as already stated. Fig.~\ref{fig0}-{\em right} provides two examples, one (dashed) in which the model is ill-defined and one in which the surface can cross both horizons and bounce. This would offer a dynamical model for a bouncing black hole. A similar configuration arises if $a_+>1$.
\end{enumerate}
This shows that, while the kinematical constraint is a necessary condition, the model has also to satisfy the dynamical constraint (\ref{conddd}) to be well defined. As an illustrative example, two possible cases are represented in the bottom figure of Figure~\ref{fig0}: the dashed line corresponds to a model in which the kinematical constraint is satisfied but the dynamical one is violated (the curves cross) while the solid blue line corresponds to a model which satisfies both constraints and is therefore well defined.

Before closing this section, let us point an interesting phenomenological implication of the generalized mass relation (\ref{Master-Equation}). Due to the quantum corrections and the (non-trivial) thin shell, the minimum energy density $\cE$ of the star is no longer fixed by its maximal radius, $R_{\text{max}}$, and its mass, contrary to the classical OS model where $M = 4\pi \rho_{\text{min}} R^3_{\text{max}}/3$ holds during the whole collapse. Now, thanks to the quantum corrections and the thin shell, the density is a free parameter of the model, adding a new dimension in its parameters space. It is therefore convenient to introduce,  as in Ref.~\cite{BenAchour:2020mgu}, the parameters
\be
\xi^2\equiv \frac{4\pi}{3}  \frac{\cE R_{\text{max}}^3}{M}\, \qquad \text{such that} \qquad R_c=\frac{\xi R_{\text{max}}^{3/2}}{\sqrt{2GM}}\,.
\ee
For the classical OS relation, one has $\xi=1$ so that this parameter encodes how dense an object can be at fixed maximal radius and mass compared to the classical case.
Notice that $\xi$ encodes the portion of the mass of the compact object which is contained in the interior region or is on the thin shell. A small $\xi$ corresponds to a compact object having most of its mass carried by the thin shell.

\subsection{Tracking the internal causal structure with the Kodama vector}

In order to follow the causal nature of the regions crossed by the thin shell, the easiest way is to compute the norm of the Kodama vector and evaluate it on the shell. Using the definition of the Kodama vector, see Refs. \cite{Kodama:1979vn, Abreu:2010ru} for details, its components read
\be
\zeta^{\alpha} \partial_{\alpha} = \left[ \epsilon^{\alpha\beta}_{\perp} \nabla_{\beta} R(\tau, \chi) \right] \Big{|}_{\chi=\chi_0}\;\partial_{\alpha} = a(\tau) \cos{\chi_0} \partial_{\tau} - \dot{a}(\tau) \sin{\chi_0} \partial_{\chi}\,,
\ee
with $\epsilon^{\alpha\beta}_{\perp} = \epsilon^{\alpha\beta}/\sqrt{\gamma}$ the densitized Levi-Civita tensor associated to the two-dimensional base space of the manifold with interior  coordinates $(\tau, \chi)$.  Its norm, related to the Misner-Sharp energy, is 
\be\label{e.kodama}
\zeta^{\alpha} \zeta_{\alpha} \big{|}_{\chi =\chi_0} = R^2(\tau) \left[ H^2(\tau) - \frac{\cot^2{\chi_0}}{R^2_c}\right]\,. 
\ee
Note that because we have used the interior metric, the norm does not depend on the parameters of the exterior geometry, namely $(M, \beta)$. A first observation to be extracted from this expression is that at the bounce, i.e. when $H=0$, the Kodama vector is time-like, i.e. $\zeta^{\alpha} \zeta_{\alpha} \big{|}_{\chi =\chi_0} <0$. This confirms that at the bounce, the shell is in an untrapped region. Moreover, the shell cannot bounce on a trapping horizon, since the norm of the Kodama vector should be zero in that case. This last observation provides an interesting difference from the construction of a black-to-white vacuum bounce in the LQG literature. Indeed, in such effective models based on the polymer regularization of the Kantowschi-Sachs interior geometry, the bounce is achieved on a spacelike-hypersurface at which both expansions vanish, i.e. $\theta_{+}(\tau_b) = \theta_{-}(\tau_b) =0$; see Refs. \cite{Bodendorfer:2019cyv, Ashtekar:2018cay} for details. However, in our framework (where the interior is modeled by the spatially closed FL cosmology), the transition is very different since at the bounce  $\theta_{+}(\tau_b) >0$ and $\theta_{-}(\tau_b) <0$ on the time-like shell. This difference comes from the fact that the Kantowski-Sachs and the closed FL universe are not isomorphic to each other, the former being anisotropic while the later is isotropic. There are therefore no reasons to expect a similar behavior of the quantum transition in these two different scenarios. Let us also emphasize that in the LQG literature, the interior of a \textit{vacuum} (eternal) black hole was modelled by a Kantowski-Sachs cosmology. Therefore, our constraints, and results, are applicable to the more realistic scenario of black hole formation from a collapsing massive object. In other words, our constraints arise when one tries to introduce quantum effects to get a black-to-white hole transition for the physically interesting matter-collapse case, as opposed to the modelling of quantum vacuum black holes.

Finally, we point that this quantity can be used to track only the interior causal structure and the formation of horizon in the $\kappa=1$ FL universe. A more detailed investigation using the exterior point of view and the matching conditions should allow to track the formation of the horizon of the exterior geometry. However, this goes beyond the scope of this work and we leave this for future investigations. 

\section{A concrete realization of black-to-white hole bounce}\label{sec3}

We can now investigate a concrete model of black-to-white hole bounce using the techniques of LQC. This model provides a generalization of our previous study~\cite{BenAchour:2020mgu} which would now also include macroscopic collapsing objects, with quantum corrections being confined to the deep interior.

\subsection{Bouncing interior geometry}
To model the interior regular geometry, we follow the Ref.~\cite{BenAchour:2020mgu} and assume it is described by spatially closed LQC~\cite{Corichi:2011pg, Dupuy:2016upu}. More precisely, we shall work in the connection regularization scheme. Following the results of Refs. \cite{Corichi:2011pg, Dupuy:2016upu}, and extended in Ref. \cite{BenAchour:2020mgu}, the dynamics is described by the modified Friedman equations
\begin{align}
\cH^2 &= \frac{8\pi G}{3} \left( \rho - \frac{1}{R^2_c a^2}\right) \left( 1- \frac{\rho -\rho_1}{\rho_c}\right)\,,\\
\dot{\cH} + \frac{1}{R^2_c a^2}& = - 4\pi G \left( \rho - \frac{2\rho_1}{3}\right)   \left( 1- \frac{\rho -\rho_2}{\rho_c}\right)\,,\\
\dot{\rho} & = - 3\cH \rho\,,
\end{align}
where $\rho_1$ and $\rho_2$ are time-dependent functions whose explicit expressions can be found in Ref.~\cite{BenAchour:2020mgu} and $\rho_c$ is a constant critical energy density given by
\be
\rho_c = \frac{3}{8\pi G \gamma^2 \tilde{\lambda}^2} = \frac{\cE}{\gamma^2 \lambda} \;, \qquad \text{with} \qquad \lambda = \frac{\tilde{\lambda}}{R_c}\,,
\ee 
where $\tilde{\lambda}$ and $\gamma$ are the UV cut-off introduced by the LQC corrections and the Barbero-Immirzi parameter, respectively, while $\cE$ is defined in Eq.~(\ref{ene}). As was already pointed out in Ref.~\cite{BenAchour:2020mgu}, the effective scale at which quantum gravity becomes non-negligible turns out be given by the dimensionless parameter $\lambda = \tilde{\lambda} /R_c$. Consequently, while the length scale $\tilde{\lambda}$ is expected to be fixed once and for all, $R_c$ changes from star to star according to its energy density and so does the scale governing quantum gravity effects in that particular star. In conclusion, the interior effective geometry is characterized by the three physical parameters $(R_{\rm max},M, \cE)$ of the collapsing star, or equivalently, by $(\chi_0,\chi_s,\xi)$ and a parameter for the quantum theory $\lambda$ since we set $\gamma=1$.

In terms of the parametrization~(\ref{para1}),  the quantum corrections are encoded in the function $\Psi_1$~\cite{BenAchour:2020mgu}, introduced in Eq.~(\ref{para1}),
\be
\label{pa}
\Psi^{\epsilon}_1 = \frac{\lambda^2}{a^2} \left( 1+ \epsilon \gamma \sqrt{\frac{1}{a} -1}\right)^2\,.
\ee
As discussed in detail in Ref.~\cite{BenAchour:2020mgu}, the dynamics exhibits two distinct minimal radii and a maximal radius. Therefore, the object experiences two consecutive cycles of collapse and expansion, starting from the maximal radius $a_{\text{max}}=1$, collapsing down to the first minima $a^{+}_{\text{min}}$, expanding again and re-collapsing to the second minima $a^{-}_{\text{min}}$ to finally reach again the maximal radius, leading to a pulsating object with a two characteristic frequencies. The existence of these two bounces is reflected in the parameter $\epsilon=\pm1$ in Eq.~(\ref{pa}). In term of the scale factor, the minimal allowed values of the scale factor scales  as
\be
a^{(\pm)}_{\text{min}}(\lambda,\gamma) =\gamma^{2/3}\lambda^{2/3}\mp{\cal O}({\lambda^2})\,,
\ee
for $\lambda \ll 1$. It is important to stress that the dynamics of the interior region is not affected by the assumed form of the outside metric so that the dynamics of the collapse, and hence $a_{\rm min}$, is identical to the one described in  Ref.~\cite{BenAchour:2020mgu}. This effective UV completion therefore provides the model of a pulsating object oscillating between $R_{\rm max}$ and $R_{\rm max}a^{\pm}_{\text{min}}$. This dynamics and its deviations compared to the standard collapsing and expanding branches of the classical spatially closed FL universe are depicted in Figure~8 of Ref.~\cite{BenAchour:2020mgu}.

\subsection{Regular exterior geometry with an inner horizon}

\label{sec4.2}
In order to investigate the role of the inner horizon, we shall work with a regular spherically symmetric black hole geometry. Indeed, imposing regularity of the interior geometry and asymptotic flatness implies that the geometry exhibits at least two horizons. Such an effective geometry provides therefore the ideal candidate to build a concrete model. Several such ad hoc geometries have been proposed so far, such as the Hayward proposal and its generalizations \cite{Hayward:2005gi, DeLorenzo:2014pta}, but most of them are quite complicated to use in practice. Moreover, while it would be interesting to work with an exterior geometry descending from effective LQG models, the available solutions do not provide both the required physical content and the simplicity needed for our model. Instead, we shall turn to another candidate metric satisfying these criteria. 

Recently, a metric describing a regular black hole with a Minkowski core was proposed by Simpson and Visser in Ref.~\cite{Simpson:2019mud}. The metric function (\ref{para}) is  given by
\be\label{e.15b}
f(r) = 1 - \frac{2m(r)}{r} = 1 - \frac{2M}{r} k\left(\frac{2GM}{r}\right) =   1 - \frac{2M }{r} \hbox{e}^{-\frac{2GM}{r}\beta}\,,
\ee
where the function $k(2GM/r)$ corresponds to the parametrization of Eq.~(\ref{e.15}) and $\beta$ is the UV cut-off coming from the exterior geometry and encodes the deviation from Schwarzschild geometry. Hence, the metric is asymptotically flat, and approximates the Schwarzschild geometry for $2GM/r \ll 1/\beta$. The new length scale $2 G M \beta$ encodes the deviation from the standard Schwarzschild geometry and can be understood as a quantum parameter descending from a UV-completion of GR in the region exterior to the collapsing star. As shown in Ref.~\cite{Simpson:2019mud}, provided that the condition~(\ref{c1}) below holds, this geometry is regular and possesses two horizons at which $f(r)=0$, i.e. at $r=-2GM\beta/W(-\beta)$, where $W$ is the Lambert function defined by $W[x]\hbox{e}^{W[x]}=x$. Since $\beta$ and $r$ are positive, it implies that $W(-\beta)$ needs to be negative, i.e. one needs to require
\be\label{c1}
0<\beta<\frac{1}{e}\,.
\ee
Then, the Lambert function enjoys two branches, the first such that $-1<W_+(-\beta)<0$ and the second such that $-\infty<W_-(-\beta)<-1$. Given the condition~(\ref{c1}), the metric has two horizons located at
\be
R_{\pm} = 2GM \hbox{e}^{W_{\pm}(-\beta)},
\ee
such that $0<R_-< 2 G M \beta < R_+<+\infty$. It corresponds to
\be
a_{\pm} = a_* \hbox{e}^{W_{\pm}(-\beta)}.
\ee
When $\beta\rightarrow0$,  we have $R_-\rightarrow 0$ and $R_+\rightarrow 2GM\equiv R_s$, which corresponds to our previous analysis~\cite{BenAchour:2020mgu}. 

\subsection{The three families of bouncing compact objects}
\begin{figure}[h]
 \centering
   \includegraphics[scale= .37]{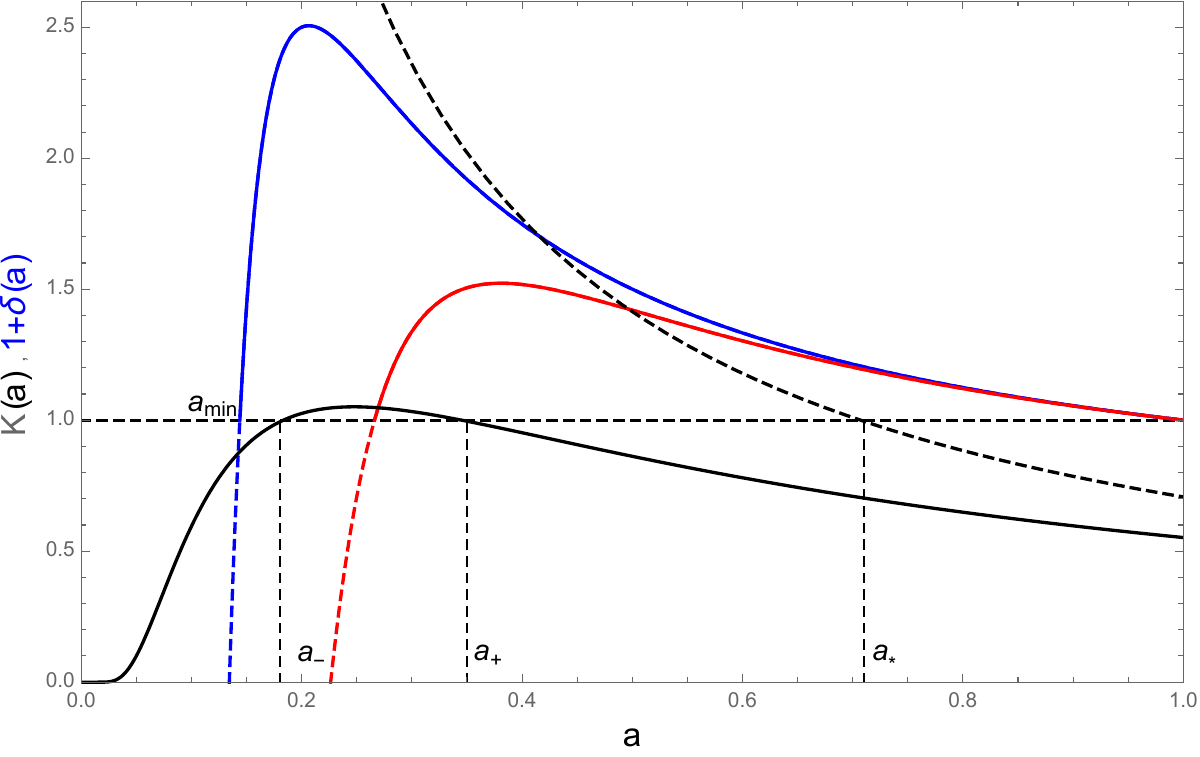} \includegraphics[scale= .37]{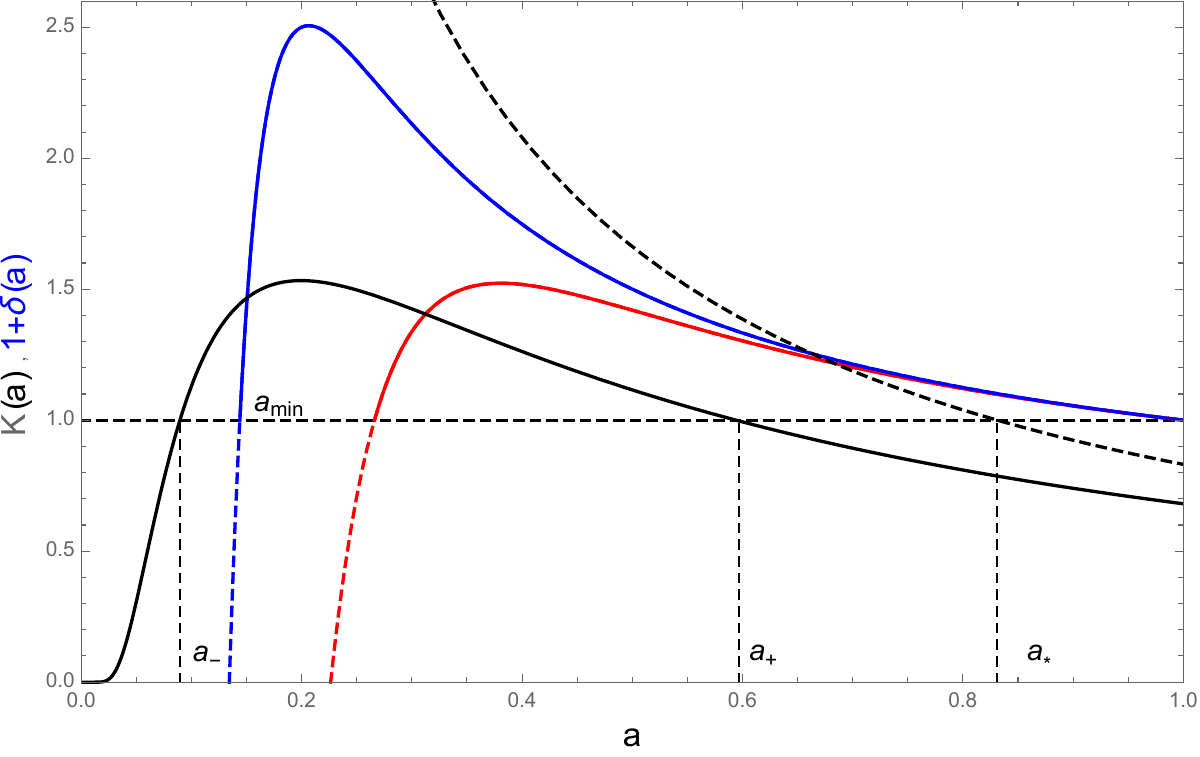}
      \includegraphics[scale= .37]{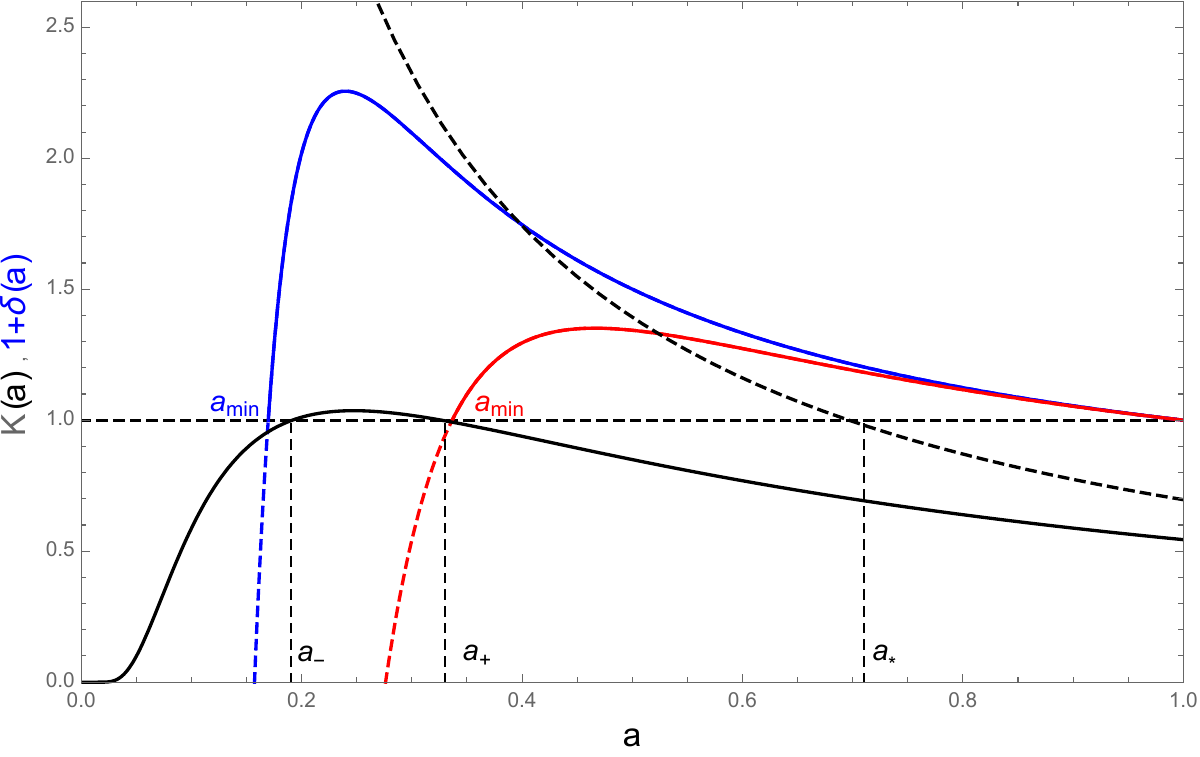} \includegraphics[scale= .37]{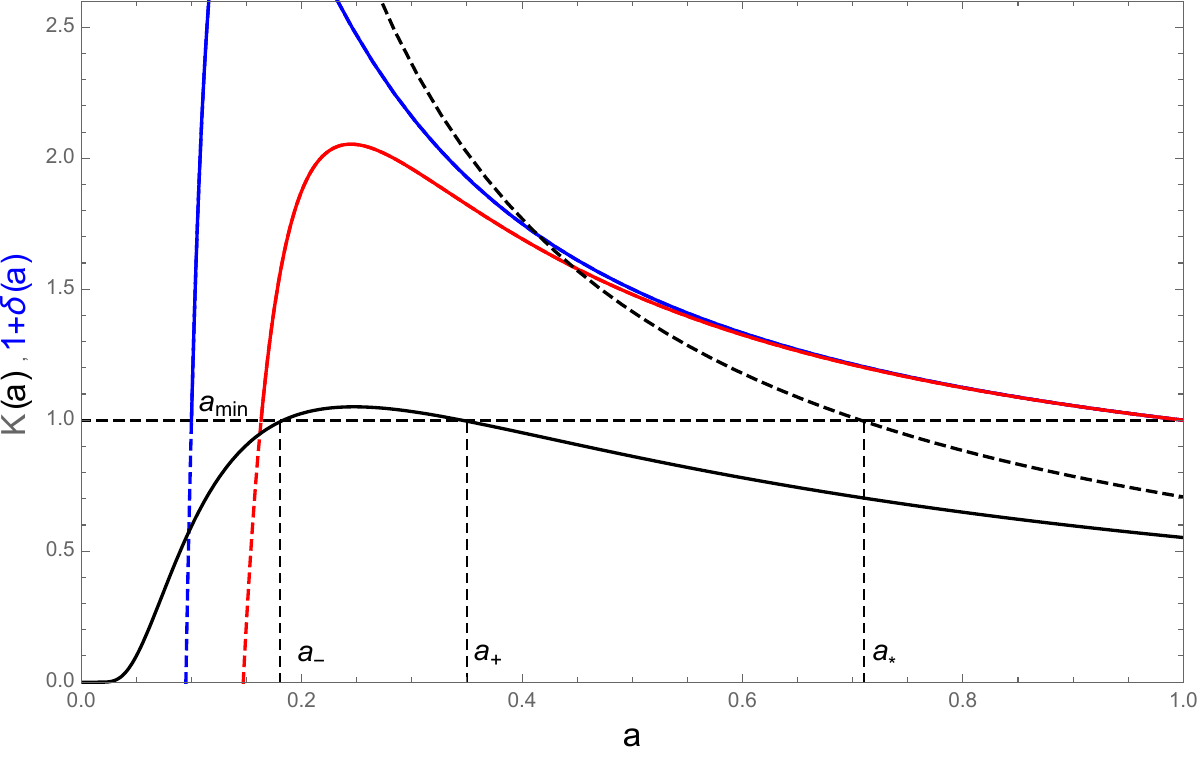}
   \caption{Bouncing black hole for an exterior geometry defined by Eq.~(\ref{e.15b}) and assuming LQC regularization. {\em Top-left}- The model is well-defined up to $a_{\rm min}<a_-$ for the minus-cycle (blue) but not for the plus-cycle (red) and thus does not provide a viable model. Parameters: $\chi_0=\pi/4$, $\chi_s=\pi/6$ for the star, $\beta=0.35$ for the external geometry and $\lambda/R_c=0.1$ for the quantum corrections. This implies $a_*=0.707$ which characterizes the radius at which the star would have crossed the horizon of a Schwarzschild black hole (dashed line). We recover graphically the condition of Ref.~\cite{BenAchour:2020mgu} that $a_{\rm min}$ needs to be larger than $a_*$ for the model to be well-defined.   {\em Top-right}- The model is ill-defined for the two cycles. Parameters: $\chi_0=\pi/4$, $\chi_s=\pi/5$, $\beta=0.24$, $\lambda/R_c=0.1$. This implies that $a_*=0.83$. {\em Bottom-left}- The model is well-defined and describes a star that is first bouncing before crossing the outer horizon (first cycle in red) and the  crosses the outer and inner horizons before bouncing (second cycle in blue). Parameters: $\chi_0=\pi/4$, $\chi_s=\pi/6$, $\beta=0.335$, $\lambda/R_c=0.14$. {\em Bottom-right}- The model is well-defined and the star crosses both horizons in its two cycles. Parameters: $\chi_0=\pi/4$, $\chi_s=\pi/6$, $\beta=0.35$, $\lambda/R_c=0.05$. All these models are indeed ill-defined if the external space is described by a Schwarzschild geometry (black dashed curve)}\label{fig2} 
\end{figure}

The issue of whether the above description provides for a model for a pulsating star experiencing black-to-white hole bounce depends on the parameters of the model. As can be seen from Fig.~\ref{fig2}, it is not sufficient that $a_{\rm min}<a_-$ since, even then the condition~(\ref{e.ccc}) can either be satisfied or not. Note also that it has to be satisfied by each of the two cycles, the two cycles being a specific characteristic of the LQC dynamics chosen for the UV-completion of the geometry describing the interior of the collapsing star. It leads to 3 classes of objects: 
\begin{enumerate}
\item[(1)] Stars that bounce above the outer horizon during their two cycles, 
\item[(2)] Stars that bounce above the outer horizon in the first cycle and then cross both horizons before bouncing in their second cycle, and 
\item[(3)] Stars that cross both horizons before bouncing during their two cycles. 
\end{enumerate}
The phenomenology is thus richer and contains bouncing stars (1), bouncing black holes (3) and a new kind of hybrid pulsating stellar objects (2) with a first cycle corresponding to a bouncing star and the second cycle to a black-to-white hole bounce.

It is easily checked that the maximum of $K(a)$ is reached in $a=\beta a_*$ and that $K(\beta a_*)=1/\beta e$. When $\beta$ decreases, $a_-$ is shifted to smaller values and $a_+$ to higher values. Similarly the peak of $\delta(a)$ increases with $\sin^2\chi_0$. The plots of Fig.~\ref{fig2} also compare the cases in which the external spacetime is described by a Schwarzschild geometry, so that, since $k(a)=1$ for that case, and for which the condition~(\ref{e.15b}) is satisfied if and only if $a_{\rm min}\geq a_*$, as was already concluded in Ref.~\cite{BenAchour:2020mgu}. 

\begin{figure}[h]
 \centering
   \includegraphics[scale= .5]{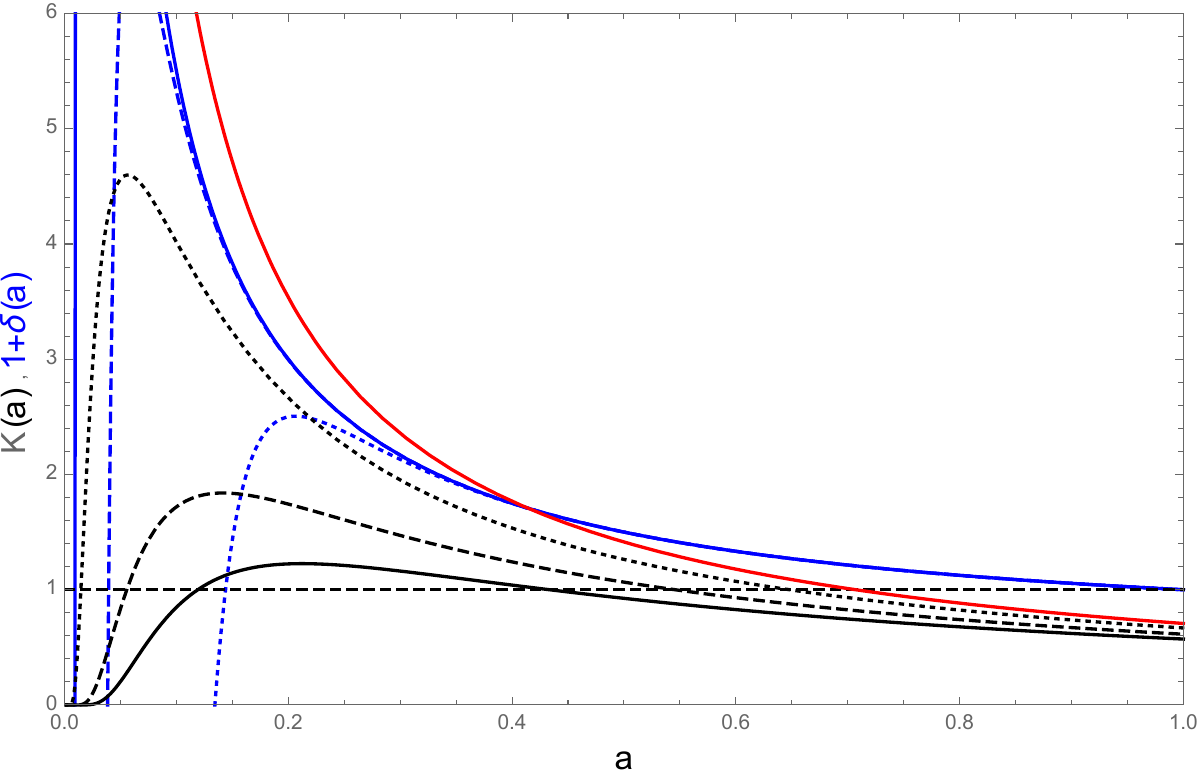}
   \caption{Dependencies of $K(a)$ (black) and $1+\delta(a)$ (blue) for $\beta=0.3$ (solid), 0.2 (dashed) and 0.1 (dotted) and $\lambda=10^{-3}$ (solid), $10^{-2}$ (dashed) and $10^{-1}$ (dotted). The red curve corresponds to $K(a)$ for a Schwarzschild external geometry.}\label{fig2b} 
\end{figure}

Figure~\ref{fig2b} illustrates the dependence on the parameters $(\lambda,\beta)$, showing that small values of $\lambda$ or with $\beta\rightarrow 1/e$ will generically lead to successful models. The case $\beta\simeq 1/e$ is particularly interesting because $a_-=a_+\simeq\beta a_*$  -- the horizons tend to merge -- and $K(a)$ is always smaller than $(\beta e)^{-1}$. If $\lambda$ is small, then the model will be well-defined as soon as, going back to physical units,
\be
\label{ine}
\frac{\tilde\lambda\gamma}{\beta^{3/2}}<2GM\xi\,,
\ee
a condition that will be easily satisfied for stellar objects since one expects $\tilde\lambda$ to be of the order of some Planck length. Finally, the different scales of the problem must satisfy the following hierarchy
\be
\tilde{\lambda} \ll  R_{-} < R_{+} < R_{\text{max}} \leqslant R_c.
\ee

\subsection{Thin-shell behavior}

Having discussed the three possible realizations of bouncing compact objects which satisfy our dynamical constraint, we turn now to the behavior of the energy and pressure of the thin shell. This shell encodes part of the quantum corrections to GR and is a  necessary ingredient of the model in order to glue the exterior and interior geometries. The dynamically allowed profiles of the surface energy $\tilde{\Sigma}$ correspond to the roots of the  second order polynomial equation (\ref{seconorder}) read
\be
\tilde{\Sigma}_{\pm} =\frac{\cot{\chi_0}}{a} \pm \frac{\sqrt{\Delta(a)}}{2}\,,
\ee
where $\Delta$ is  given by Eq.~(\ref{conddd}) and $\tilde{\Sigma}$ has been defined below Eq.~(\ref{seconorder}). We see that the first branch $\tilde{\Sigma}_{+}$ will always remain positive while the second branch $\tilde{\Sigma}_{-}$ can experience negative values.
At this stage, both branches can be considered as admissible. In order to choose which one should be physically favored, let us consider their behavior in the classical limit and near the bounce.

The classical limit of this model corresponds to an extension of the Oppenheimer-Snyder model which includes a non-vanishing thin shell. In term of the parameter of the model, this classical regime is obtained by sending $\beta \rightarrow 0$ and $\lambda/R_c \rightarrow +\infty$ such that both quantum corrections in the exterior and in the interior geometries vanish. In that classical regime, the junction condition still allow for a non vanishing thin shell energy with two possible branches. It is instructive to note that the standard Oppenheimer-Snyder model is recovered in this regime provided one chooses the branch $\tilde{\Sigma}^{\text{GR}}_{-}$ and set additionally $\sin{\chi_s} = \sin^3{\chi_0}$. In that case, one has $\tilde{\Sigma}^{\text{GR}}_{-} =0$. At the end of the day, the classical limit of our model can be understood as an extension of the GR solution represented by the seminal Oppenheimer-Snyder model supplemented with a thin shell.

\begin{figure}[h]
 \centering
   \includegraphics[scale= .55]{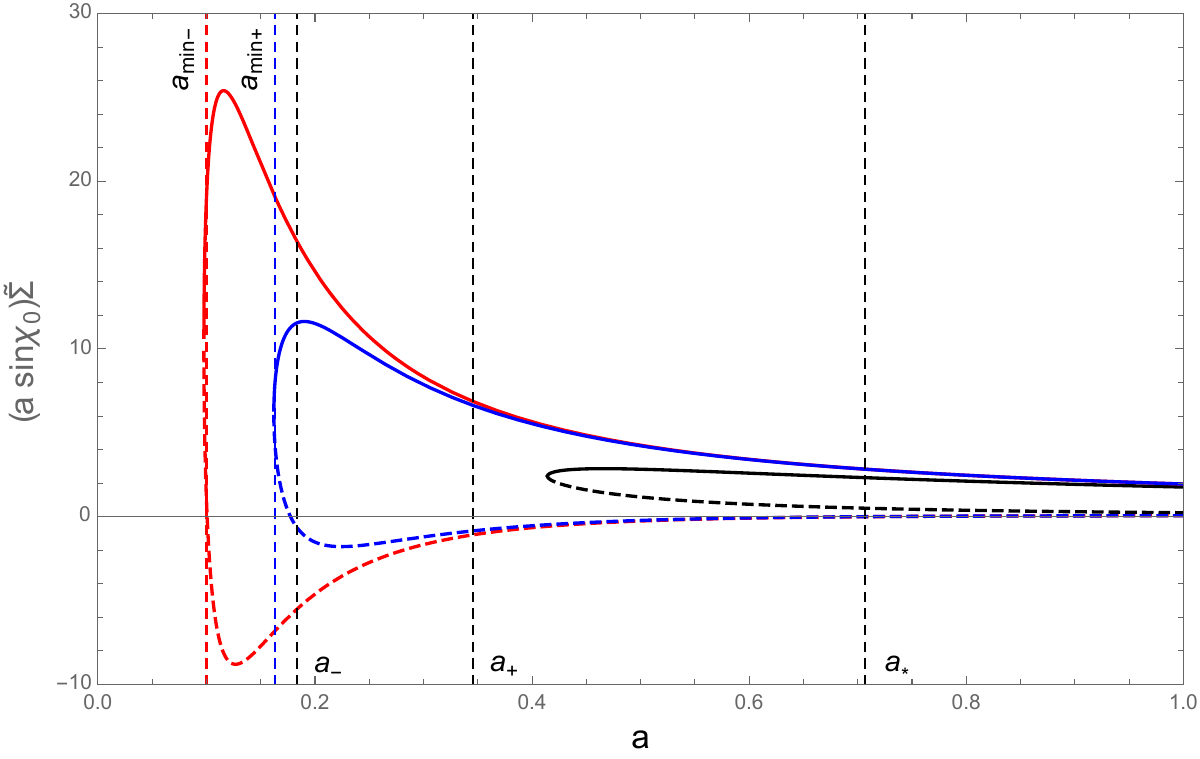}
   \includegraphics[scale= .55]{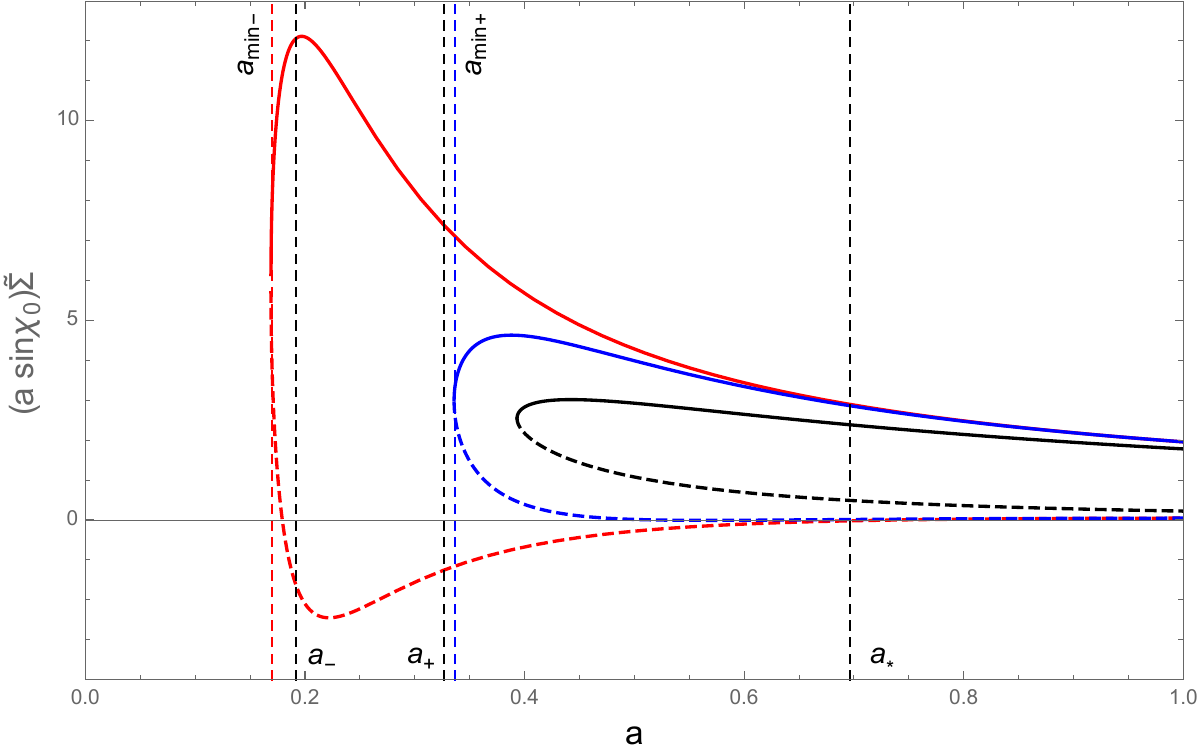}
   \caption{Rescaled surface energy $\tilde{\Sigma}(a)$ of the thin shell during each phases as a function of the scale factor $a(\tau)$. We have chosen $G=1$ and $\chi_0 = \Pi/4$. \textit{Up} : Two consecutive black-to-white hole bounce (blue and red curves) and its classical counterparts (black curve). The other parameters are fixed to: $\chi_s = \pi/6$, $\beta = 0.35$, $\lambda = 0.05$. \textit{Bottom}:  Hybrid case with one bouncing star without horizon formation (blue) and one black-to-white hole bounce with outer and inner horizon formation (red) and its classical counterparts (black). The other parameters are fixed to $\chi_s = \pi/(6.1)$, $\beta = 0.355$, $\lambda = 0.14$. The solid curves correspond to the (rescaled) positive branch $\tilde{\Sigma}_{+}$ and the dotted  curves to the second branch $\tilde{\Sigma}_{-}$.}\label{figSigmae} 
\end{figure}

Consequently, the quantum extension presented in this work implies that the effective thin shell contains both classical as well as quantum contributions. A crucial requirement is that the modified effective thin shell approaches its classical GR counterpart when the quantum corrections vanish.  The two different branches $\tilde{\Sigma}_{\pm}$ as well as their classical counterpart $\tilde{\Sigma}^{\text{GR}}_{\pm}$ are depicted in Figure~\ref{figSigmae} below in two different cases. The first case on the top figure corresponds to profiles of the effective thin shell during the two cycles of a black-to-white hole bounces (the blue and red curves) while the GR solution is depicted in black. Solid line corresponds to the first branch while the dotted lines correspond to the second branch. The second case on the bottom figure corresponds to a hybrid object with one horizonless cycle followed by a black-to-white hole bounce. It is straightforward to see that both branches $\tilde{\Sigma}_{\pm}$ approach their classical counterpart in the regime where $a\rightarrow 1$, i.e in the classical regime. So from the point of view of the classical limit, both branches satisfy the standard requirement. Let us further point that if we impose that for $\sin{\chi_s} = \sin^3{\chi_0}$,  the shell has to vanish in order to reproduce the seminal Oppenheimer-Snyder model, then only the branch $\tilde{\Sigma}_{-}$ can be selected.
Finally, a last possible  requirement might be motivated by the fact that the thin shell encodes part of the quantum effect. As such, it might be interesting to select the branch which violates the weak energy condition during the process. Again, this would select the branch $\tilde{\Sigma}_{-}$ which is the only one that can have negative values. Finally, let us point that the trace of the surface energy momentum tensor given by Eq~(\ref{press}) is independent of the chosen branch for $\tilde{\Sigma}$. This concludes the discussion on the behavior of the effective thin shell.

\section{ Discussion} 
Using the thin shell formalism, we have presented a new model-independent constraint for bouncing black hole geometries. The first argument (Section~\ref{sec1}) is purely kinematical and thus valid for the gluing of {\em any} spherically symmetric geometries across a time-like thin shell provided a bounce occurs in the interior region, the time-like thin shell being placed at the surface of the collapsing matter. This general argument leads to the constraint (\ref{condd}) which sets a bound on the minimal allowed radius of the shell at the bounce, encoded in the exterior Misner-Sharp mass function. The general consequence of this bound is that the collapsing object can bounce only when the thin shell is in an untrapped region or, precisely on a trapping horizon, a property which is made explicit by the constraint (\ref{e.14}) on the expansions. We emphasize that this conclusion is rather general, and only assumes {\em i}) spherical symmetry, {\em ii}) a time-like shell as the matching hypersurface (a physical requirement for any realistic collapsing star), and finally {\em iii}) a singularity-free bouncing interior geometry. The purely kinematical nature of this bound is made explicit by the fact that we never used the second junction condition involving the extrinsic curvature but only the continuity of the induced metric. Therefore, this kinematical constraint allows one to extract a very general property of such bouncing compact objects which is completely model-independent: namely,\textit{ that the bounce can only occur either outside the outer horizon or inside the inner horizon}. This is our main result. 

However, a key point is that this kinematical constraint is a necessary but not a sufficient condition to model a \textit{consistent} black-to-white hole bounce. To go further, one has to make explicit choices for the dynamics. Therefore, in the second part of this work, we have investigated the consequences of this constraint on a specific gluing between an exterior static geometry and a bouncing spatially closed FL cosmology with dust.  We have introduced a general parameterization of the Misner-Sharp mass function and the quantum corrections to the Friedman equations in order for the model to remain general enough. In order to glue these two geometries, we have assumed that the second junction conditions take the form of the Israel-Darmois matching constraints. Although they correspond to the classical ones, we allow the perfect fluid sourcing the discontinuity of the extrinsic curvature to violate the standard energy conditions in a local region where quantum gravity terms should dominate. As such, the energy content of the thin shell encodes part of the quantum effects. Solving the junction conditions, we have obtained a generalized mass relation (\ref{Master-Equation}) relating the exterior Misner-Sharp mass function, the surface energy of the shell and the quantum corrections. This matching condition has to remain satisfied during the whole collapse. Interestingly, this constraint provides an efficient way to select a well-defined bouncing dynamics through the positivity of the determinant (\ref{conddd}). 

As expected from the kinematical constraint, we have shown that depending on the position of the minimal radius $R_{\text{min}}$ w.r.t. the positions of the inner and outer horizons, $R_{\pm}$, different family of objects can be identified when imposing the above consistency conditions. When $ R_{+} < R_{\text{min}}$, the collapsing object never forms a trapped region and leads to a pulsating horizonless star. When $ R_{-} < R_{\text{min}} < R_{+}$, the model is not well-defined. Finally, black-to-white hole bounce can be realized only when $ R_{\text{min}} < R_{-}$. Interestingly, the pictorial representation of this dynamical constraint (see Fig.~\ref{fig0}) provides an efficient and clean way to determine whether a given model will be well-defined or not, depending on one hand on the parameters, and on the other hand on the functions $\left( k(a), \Psi_1(a)\right)$. 

In the end, we have proposed a concrete realization of a black-to-white hole bounce satisfying the derived consistency constraints. We have described the exterior spacetime by the regular black hole metric of Ref. \cite{Simpson:2019mud} and the interior spacetime by the spatially closed LQC bouncing cosmology~\cite{Corichi:2011pg}. This extends the conclusions of our previous work \cite{BenAchour:2020mgu}, in different directions.
\begin{itemize}
\item First, since the bounce can now occur well inside the black hole, at a radius smaller than the inner horizon, quantum effects can remain confined to the deep interior. The weak energy condition turns out to be violated near the bouncing point and, as such, the inner horizon allows one to keep this quantum effects ``hidden''\footnote{Note, however, that the outer trapping horizon in this case is not the event horizon and the interior of the inner horizon can be ``seen'' by observers outside the horizon in principle.}. Removing the inner horizon (i.e. the limit $\beta\rightarrow 0$) one would end up with a Schwarzschild exterior and a violation of the WEC outside the Schwarzschild radius as discussed in Ref. \cite{BenAchour:2020mgu}.
\item Second, the presence of an inner horizon also modifies the range of applicability of the model. As shown in Ref. \cite{BenAchour:2020mgu}, a single horizon exterior geometry imposes drastic conditions on the parameters of the model and only Planckian relics were then possible to be modeled. However, in the present case, the inner horizon leads to a new constraint on the parameters of the model, see Eq.~(\ref{ine}). This condition can now in principle allow for the description of macroscopic (i.e. stellar) objects.  However, whether macroscopic object can experience such bounce and therefore whether coherent quantum gravity effects can extend to macroscopic scales can only be answered within a full quantum theory of gravity. Condensation mechanism of quantum geometry d.o.f as discussed in the context of GFT condensates~\cite{Oriti:2018qty, Oriti:2016ueo, Gielen:2013naa} might provide an interesting way to realize this. 
\item Third, as already shown in details in Ref. \cite{BenAchour:2020mgu}, this bouncing interior geometry exhibits two distinct bouncing points $R^{\pm}_{\text{min}}$ and thus two consecutive phases of collapse/expansion. Consequently, the class of objects becomes {\it phenomenologically} richer depending on the position of the bouncing points $R^{\pm}_{\text{min}}$ w.r.t. the inner and outer horizons $R_{\pm}$. Besides objects behaving as bouncing stars or bouncing black holes in both phases, we have identified new hybrid objects alternating bouncing star and bouncing black hole behaviors. The realization of these different cases depends on the values of the exterior and interior UV cut-off $(\lambda,  \beta)$. However, notice that this property is a peculiarity descending from the choice of the interior geometry described by LQC dynamics. Nevertheless, it suggests an interesting phenomenology for such bouncing compact objects.
\end{itemize}
This minimal model of a black-to-white hole bounce, therefore, opens several perspectives and should serve a platform for further developments. Let us nevertheless point out several limitations of the present construction. Firstly, we have used the standard Israel-Darmois junction conditions to perform our gluing. In principle, the form of the junction conditions to be used should descend from the UV theory under consideration. Since this derivation is out of reach at the moment, the choice of working with the ID conditions is necessary in order to make progress. Nevertheless, it is also worth keeping in mind that the dynamics of the shell strongly depends on this simplifying choice. However, one can also argue that the two geometries to be glued can be viewed as solutions of GR with a non-trivial (and possibly exotic) energy-momentum tensor. Interpreting the two solutions considered here in that way, one can consistently investigate their gluing as done in this work. Therefore, even if our gluing appears rather artificial, it should be viewed only as a preliminary attempt to explore model-independent consistency conditions for such bouncing compact objects. In a second step, it would be interesting to consider instead two geometries descending from the same theory. Having listed the main limitations of the construction, let us now mention several interesting directions it opens.

A crucial question to be investigated will be the characterization of the bouncing time of these objects and how it scales with the black hole mass. Since it can include stellar mass objects, the astrophysical signatures of these bouncing stars/black-hole need to be investigated, along the same lines of Refs.~\cite{Barrau:2014hda, Barrau:2015uca, Barrau:2016fcg, Barrau:2018kyv, Rovelli:2017zoa}. Moreover, a careful study of the formation and development of the trapping horizons associated to this bouncing compact object would also be of interest, following for example the recent investigation of Ref. \cite{Chatterjee:2020khj}. We postpone these interesting questions to future works. 

Another crucial point would be to understand how the kinematical constraint (\ref{condd}) translates at the quantum level. Different approaches can be used to that end, using either canonical methods such as, e.g., in Refs.~\cite{Piechocki:2020bfo, Schmitz:2019jct}, the techniques of quantum reduced loop gravity~\cite{Alesci:2018loi} or the systematical approach used for example in Refs. \cite{Bojowald:2010qm, Brizuela:2014cfa, Alonso-Serrano:2020szo, Bojowald:2015fla}. This last method provides a systematic way to derive moment corrections in the effective equations satisfied by the expectation values of quantum operators, thus going beyond effective models. This might be applied to investigate the possible deviations to the Eq~(\ref{condd}) induced by the quantum fluctuations of the geometry. 

Before concluding, let us point that while the formation of an inner horizon appears to be a key ingredient of the present model, it might also generate new difficulties. Indeed, it is well-known that such structures typically generate pathological instabilities known as mass inflation; see e.g. Refs.~\cite{Eardley:1974zz, Carballo-Rubio:2018pmi, Brown:2011tv}. The same is true of the white hole horizon, which is also known to be generically unstable. Adding dynamical evaporative effects might produce geometries without Cauchy horizons, but that this is outside the scope of this work. More explicit investigations are required to come to a definitive conclusion about these issues.

We conclude by emphasizing that the major outcome of this work is to show that even if a full quantum theory of gravity is not yet available, some knowledge and astrophysical phenomenology can still be gained by considering general model-independent arguments such as the kinematical constraint presented in the first section. This is an encouraging step towards a better understanding of non-singular gravitational collapse.

\acknowledgments
The authors would like to thank Francesco Di Filippo, Edward Wilson-Ewing, Martin Bojowald, Hongguang Liu and Daniele Pranzetti for their feedbacks on the first version of this draft. The work of JBA was supported by Japan Society for the Promotion of Science Grants-in-Aid for Scientific Research No. 17H02890. SB is supported in part by funds from NSERC, from the Canada Research Chair program, by a McGill Space Institute fellowship and by a generous gift from John Greig. The work of SM was supported by Japan Society for the Promotion of Science Grants-in-Aid for Scientific Research No. 17H02890, No. 17H06359, and by World Premier International Research Center Initiative, MEXT, Japan.

\appendix
\section{Robustness of the kinematical constraint against gauge choices}

The Gullstrand-Painlev\'e gauge, for a general spherically symmetric spacetime, takes the form
\begin{eqnarray}\label{GP}
\dd s^2_{\pm} &= - e^{2\Phi(T_{\pm},R_{\pm})} \left(1 - \frac{2M_{\pm}(T_{\pm},R_{\pm} )}{R_{\pm}}\right) \dd T_{\pm}^2  + 2 e^{\Phi(T_{\pm},R_{\pm})} \frac{2M_{\pm}(T_{\pm}, R_{\pm})}{R_{\pm}} \dd T_{\pm} \dd R_{\pm} \nn\\
& \qquad \qquad \qquad \qquad \;\; + \;\dd R_{\pm}^2 + R_{\pm}^2 \dd\Omega^2\;\;\;\;\;\;\;\;
\end{eqnarray}
Proceeding as before, it is not difficult to see that the continuity of the induced metric on the time-like shell leads to the two conditions
\begin{eqnarray}\label{GP_keycond1}
& e^{2\Phi(T_{\pm},R_{\pm})} \left(1 - \frac{2M_{\pm}(T_{\pm},R_{\pm})}{R_{\pm}}\right)\dot{T}_{\pm}^2   -  2\, e^{\Phi(T_{\pm},R_{\pm})} \frac{2M_{\pm}(T_{\pm},R_{\pm})}{R_{\pm}} \dot{T}_{\pm} \dot{R}_{\pm}  - \dot{R}_{\pm}^2 = 1\,.\\
\label{GP_keycond2}
& R_{+} = R_{-}
\end{eqnarray}
where a dot corresponds to a derivative w.r.t. the proper time of the shell. These conditions are the counterpart of Eqs~(\ref{keycond}) for the free-falling observer.
At the bounce point, where the radius of the shell vanishes, i.e. $\dot{R}(\tau) \big{|}_{\tau_{\ast}}=0$, we get from (\ref{GP_keycond1}) that
\begin{eqnarray}\label{GP_A}
A^2 \big{|}_{\tau_{\ast}} = e^{2\Phi_{+}(T_{+}(\tau_*),R(\tau_*))}\left(1 - \frac{2M_{+}(T_{+}(\tau_*),R(\tau_*))}{R(\tau_*)}\right) > 0 \;\;
\end{eqnarray}
for the lapse function associated to an observer on the shell with the exterior time coordinate $T_{+}$. This coincides with Eq~(\ref{atb}) derived in Section~\ref{sec1}.
Thus, it seems that our kinematical constraint holds for gauges for which there exists no coordinate singularity at the horizon, and as such, is a robust physical result and not an artifact of a specific coordinate choice.

\end{document}